\documentclass{jfm}
\usepackage{fancyhdr}
\usepackage{amsmath}
\usepackage{amssymb}
\usepackage{graphicx}
\usepackage{natbib}
\usepackage{psfrag}
\usepackage{lscape}
\usepackage{arydshln}
\usepackage{longtable}
 \usepackage{setspace}
 \usepackage{color}
 \newcommand{\revadd}[1]{#1}

\newcommand{\Da}{\mathrm{Da\,}}
\newcommand{\St}{\mathcal{S}}

\newcommand{\Rm}{\mathrm{Rm}}
\newcommand{\Rmo}{\mathrm{Rm}_0}

\newcommand{\Cc}{\mathcal{C}}

\newcommand{\partiald}[2]{\frac{\partial #1}{\partial #2}} % command to do partial derivatives for display
\newcommand{\uvec}{\mathbf{u}} % commandfor vector velocity u
\newcommand{\nvec}{\mathbf{n}} % commandfor vector velocity n
\begin{document}
%\markboth{\footnotesize \textit{Preprint: for limited circulation only}}{\footnotesize \textit{Preprint: for limited circulation only}}
%\pagestyle{myheadings}

% JFM Title
 \title[Mushy layer convection with chimneys]{Nonlinear mushy-layer convection with chimneys: stability and optimal solute fluxes} 
% \title[Optimal solute fluxes from mushy layers]
%{Optimal solute fluxes from mushy layers} 
\author[A. J. Wells, J. S. Wettlaufer  \& S. A. Orszag ]
{A\ls N\ls D\ls R\ls E\ls  W\ns J.\ns W\ls E\ls L\ls L\ls S$^{1,3}$,% 
\ns J.\ls S.\ns W\ls E\ls T\ls T\ls L\ls A\ls U\ls F\ls E\ls R$^{1,2,3,4}$ % 
\ns \\ \and S\ls T\ls E\ls V\ls E\ls  N\ns A.\ns O\ls R\ls S\ls Z\ls A\ls G$^{3}$\footnote{\noindent Deceased 1st May 2011}}
%\author{A. J. Wells$^{1,3}$, J. S. Wettlaufer$^{1,2,3}$   \& S. A. Orszag$^{3}$ \\ 
\footnotesize
\affiliation{$^{1}$Department of Geology and Geophysics, Yale University, New Haven, CT, 06520, USA\\ \footnotesize$^{2}$Department of Physics, Yale University, New Haven, CT, 06520-8109, USA\\ \footnotesize$^{3}$Program in Applied Mathematics, Yale University, New Haven, CT, 06520, USA\\ \footnotesize$^{4}$ NORDITA, Roslagstullsbacken 23, SE-10691 Stockholm, Sweden}

 \maketitle

\begin{abstract}
We model buoyancy-driven convection with chimneys -- channels of zero solid fraction -- in a mushy layer formed during directional solidification of a binary alloy in two dimensions. A large suite of numerical simulations is combined with scaling analysis in order to study the parametric dependence of the flow. Stability boundaries are calculated for states of finite-amplitude convection with chimneys, which for a narrow domain can be interpreted in terms of a modified Rayleigh number criterion based on the domain width and mushy-layer permeability. For solidification in a wide domain with multiple chimneys, it has previously been hypothesized that the chimney spacing will adjust to optimize the rate of removal of potential energy from the system. For a wide variety of initial liquid concentration conditions, we consider the detailed flow structure in this optimal state and derive scaling laws for how the flow evolves as the strength of convection increases. For moderate mushy-layer Rayleigh numbers these flow properties support a solute flux that increases linearly with Rayleigh number. This behaviour does not persist indefinitely, however, with porosity-dependent flow saturation resulting in sub-linear growth of the solute flux for sufficiently large Rayleigh numbers. Finally, we consider the influence of the porosity dependence of permeability, with a cubic function and a Carman-Kozeny permeability yielding qualitatively similar system dynamics and flow profiles for the optimal states.

 \end{abstract}

\section{Introduction\label{sec:intro}}

Solidifying binary alloys occur in a wide variety of geophysical, geological and industrial settings. In many contexts a \emph{mushy layer} is formed as a result of morphological instability of an advancing solid-liquid interface. A mushy layer consists of a reactive porous medium of solid dendrites in local thermodynamic equilibrium with the interstitial   concentrated fluid~\cite[e.g.,][]{Worster:2000}. An important geophysical example is the growth of sea ice via the solidification of salty water~\cite[][]{Feltham:2006gf}, with $\sim 10^6$-$10^7$ square kilometres of the Arctic and Antarctic oceans freezing and melting each year~\citep{Comiso:2010}. The drainage of brine from growing sea ice modifies   the thermal and mechanical properties of the ice~\citep[see e.g.,][for a recent review]{PetrichEicken:2010} and plays a significant role in the formation of dense water masses that contribute to the ocean thermohaline circulation~\citep[see e.g.,][for a recent review]{Brandonetal:2010}. In industrial settings, defects in metal castings can arise from compositional heterogeneities~\citep{Copley:1970qt}, and so it is important to understand the evolution of solidifying alloys. Here we focus on the influence of buoyancy-driven convection on the drainage of solute from a growing two-dimensional mushy layer.

Buoyancy-driven convection of the interstitial fluid in a mushy layer occurs when the density gradient is convectively unstable~\cite[see][for a review]{Worster:1997}. The convective flow transports interstitial fluid into regions of differing concentration, which results in either local dissolution or growth of the solid matrix in order to maintain the composition in a state of local thermodynamic equilibrium. In regions where dissolution decreases the solid fraction, the permeability is increased leading to flow focussing. The nonlinear development of this convective-flow instability forms drainage channels of zero solid fraction, or \emph{chimneys}, through which buoyant plumes drain from the mushy layer into the neighbouring liquid. Experimental observations are consistent with chimney formation occurring for sufficiently large values of an appropriate mushy-layer Rayleigh number, which characterizes the strength of buoyancy compared to viscous and thermal dissipation. For directional solidification, where a material sample is translated between hot and cold heat exchangers,  chimneys form for sufficiently small solidification rates $V$ so that the resulting steady-state mushy layer thickness $h$ is large enough to generate a supercritical Rayleigh number~\citep[e.g.,][]{Peppinetal:2008,Whiteoak:2008kx}. \cite{Zhongetal:2012} showed that  the horizontal wavelength provides a further constraint on chimney formation in cells of finite width. Chimney formation also occurs after the mushy-layer thickness $h$ exceeds a critical value during transient solidification from a fixed cold boundary~\cite[see][for a review]{Aussillousetal:2006}. Transient growth experiments also indicate that the pattern of chimneys continues to evolve over time, with the extinction of flow in certain chimneys leading to an increase in the mean spacing between chimneys as the mushy layer grows thicker~\cite[][]{Wettlaufer:1997rz, Solomon:1998rc}. 

Theoretical studies have identified critical Rayleigh numbers for the onset of convection under a wide range of growth conditions by applying linear and weakly-nonlinear stability analyses \cite[see][for a review and a more recent summary]{Worster:1997,GUBA:2010rz}.  The formation of chimneys presents additional challenges for modelling, which must describe and resolve the change in flow dynamics between the interior of the porous mushy layer and the purely liquid region in the chimney. Single-domain methods, such as the enthalpy method approach of  \cite{Katz:2008tg}, have simulated solidification with small numbers of chimney-like features in a variety of configurations~\cite[see][for a summary]{Zhongetal:2012}. In particular, the simulations of  \cite{Katz:2008tg} for directional solidification in a quasi-two-dimensional Hele-Shaw cell suggest that the mean chimney spacing scales in proportion to the depth of the mushy layer. An alternative approach takes advantage of the fact that chimneys are narrow compared to their depth thereby allowing the use of lubrication theory to parametrize the flow in the chimney. \cite{Worster:1991} considered axisymmetric flow with an isolated chimney, whilst \cite{SchulzeWorster:1998} and~\cite{ChungWorster:2002} developed a model for two-dimensional steady flow with a periodic array of chimneys with imposed spacing. Theories that require the unknown chimney spacing to be imposed in advance of the calculation have several advantages and varying degrees of sophistication.  For example, \cite{Wellsetal:2010} combined this constraint with a hypothesis of optimal potential energy fluxes, so that the system selects the chimney spacing that yields the maximal rate of drainage of potential energy to efficiently drive the system towards thermodynamic equilibrium. This hypothesis predicts chimney spacings proportional to mushy layer depth and a solute flux that increases linearly with Rayleigh number over the simulated range. The resulting solute flux scaling predicts initial desalinization rates during sea ice growth that are consistent with experimental observations~\citep{Wellsetal:2011}.  An imposed chimney spacing approach also allows the development of an asymptotically-reduced model that reveals the flow structure for mushy layers of constant permeability  and assumed constant background temperature gradient~\citep{ReesJonesWorster:submit}. 

We here explore in detail the framework of \cite{Wellsetal:2010} to determine the leading order structure of the flow and its parametric dependence for differing system properties. We consider convection during the directional solidification of a two-dimensional mushy layer with a periodic array of chimneys. In \S\ref{sec:model} we describe the theoretical model and numerical methods employed. In \S\ref{sec:lambdaresults} a broad array of numerical simulations are used to demonstrate the evolution of the system dynamics and stability as the chimney spacing and vigour of convection change. We then investigate the dynamics of states with optimal potential energy flux in \S\ref{sec:optimalresults}, and for a variety of liquid concentrations determine asymptotic scaling laws for flow properties as the strength of convection increases. We consider the influence of the permeability variation with solid fraction in section \S\ref{sec:permresults}, before concluding in \S\ref{sec:discussion}.

\section{A model of non-linear convection with chimneys\label{sec:model}}

\begin{figure}
\centering
\psfrag{LIQUID}{\textsc{Liquid}}
\psfrag{MUSH}{\textsc{Mush}}
\psfrag{SOLID}{\textsc{Solid}}
\psfrag{CHIMNEY}{\textsc{Chimney}}
\psfrag{L}{$l$}
\psfrag{V}{$V$}
%\psfrag{azt}{$2 a(z,t)$}
\psfrag{azt}{$2 a$}
\psfrag{zeq0}{$z=0$}
\psfrag{zeqhxt}{$z=h(x,t)$}
%\psfrag{TCPhiPsi}{$T$, $C$, $\phi$, $\hat{\mathbf{u}}$}
\psfrag{u}{}
\psfrag{CeqCETeqTE}{$C=C_E,\quad T=T_E$}
\psfrag{CeqCoTeqTo}{$C=C_0,\quad T=T_{\infty}$}
\includegraphics[height=4.5cm]{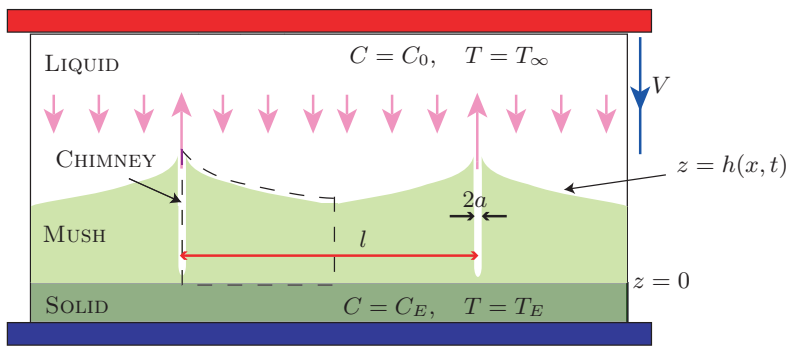}
\caption{Schematic illustration of the model configuration for directional solidification. Liquid of far-field salinity $C_0$ and temperature $T_{
\infty}$ is translated at velocity $V$ between hot and cold heat exchangers, and partially solidifies to form a mushy region of thickness $h(x,t)$. A eutectic solid forms at the lower heat exchanger with temperature $T_E$ and concentration $C_E$ at the mush-solid interface. In this geometry, with gravitational acceleration $-g\mathbf{k}$ aligned perpendicular to the heat exchangers, solidification releases buoyant fluid of relative density $\rho_o g \beta(C-C_E)$ to drive the convective flow, where $\rho_o$ is a constant reference density and $\beta$ the constant haline coefficient. 
We consider a periodic array of chimneys of dimensional width $2\hat{a}(z,t)$ and imposed spacing $l$, with the dashed outline indicating the simulation region.
Buoyant solute-depleted plumes exit the mushy layer via chimneys, and there is a weak return flow in the overlying liquid as indicated by arrows in the liquid region. Along with other systems, this geometry is consistent with the solidification of a trans-eutectic aqueous ammonium chloride solution cooled from below, where cooling at the base of the mushy layer lowers the liquidus temperature and results in lower solute concentration of the interstitial fluid. 
}
\label{fig:notation}
\end{figure}

Building on the previous models of \cite{SchulzeWorster:1998} and~\cite{ChungWorster:2002}, we apply the framework of~\cite{Wellsetal:2010} to describe  two-dimensional directional solidification, as illustrated in figure~\ref{fig:notation}. 
A two-component mixture of liquid concentration $C$ and temperature $T$ is translated at a velocity $V$ between hot and cold heat exchangers, and forms a mushy region of thickness $h(x,t)$.  In the illustrated setting, solidification within the mushy region lowers the local concentration of the liquid, and hence provides the buoyancy to drive convection. This would occur in aqueous $\mathrm{NH}_4 \mathrm{Cl}$ solidified from below and the dynamics and thermodynamics are ostensibly the same as ice forming above a region of salt water. We investigate the behaviour of a periodic array of chimneys within this system.  For a given chimney spacing $l$, we calculate the resulting solute flux.

The thermodynamic evolution of the mushy region is treated within the framework of so-called ideal mushy layer theory~\cite[for example, see][for a full discussion of the underlying physics]{Worster:2000,SchulzeWorster:2005}. It is assumed that the specific heat capacity $c_p$ and thermal diffusivity $\kappa$ are constant across both solid and liquid phases, the liquid has kinematic viscosity $\nu$, and $L$ is the latent heat of fusion. Following~\cite{SchulzeWorster:1998}, we scale velocities by $V$, and length and timescales via the thermal diffusion scales $\kappa/V$ and $\kappa/V^2$ respectively. The temperature and concentration within the mushy layer are constrained by the condition of local thermodynamic equilibrium, with the liquid solute concentration maintaining a state of local phase equilibrium by either dissolving or crystallizing the solid matrix faster than any timescale of macroscopic transport.
Hence the liquid concentration is slaved to the temperature via the liquidus curve $T=T_L(C)=T_E+\Gamma(C-C_E)$, where $\Gamma$ is constant.  Both temperature and liquid concentration can then be characterised by the local dimensionless temperature 
\begin{equation} \theta = \frac{T-T_L(C_0)}{T_L(C_0)-T_E}=\frac{C-C_0}{C_0-C_E}, \label{eq:thetadefn} \end{equation}
where $C_0$ is the concentration in the bulk liquid layer (see Fig.~\ref{fig:notation}). The dimensionless temperature $\theta$, solid fraction $\phi$, Darcy velocity $\uvec$ and pressure $p$ then satisfy 
 \begin{subeqnarray} \gdef\thesubequation{\theequation \textit{a,b}}
 \uvec = -\Rm \,\Pi \left(\nabla p +\theta \mathbf{k}\right), \qquad \qquad \nabla \cdot \uvec &=0, \label{eq:Darcyu}  \end{subeqnarray}  \returnthesubequation
 \begin{align}
\partiald{\theta}{t}-\partiald{\theta}{z}+\uvec \cdot \nabla \theta&= \nabla^2 \theta +\St\left(\partiald{\phi}{t}-\partiald{\phi}{z}\right), \label{eq:heat} \\
\left(\partiald{}{t}-\partiald{}{z}\right)\left[ (1-\phi)\theta+\Cc \phi\right] +\uvec \cdot \nabla \theta &= 0, \label{eq:salt}
\end{align}
representing conservation of momentum, mass, energy and solute.
We have assumed that the fluid density depends linearly on concentration, that solute diffusion is negligible compared to thermal transport, and that the mushy layer has a porosity-dependent dimensional permeability $\hat{\Pi}=\Pi_0\Pi(\phi)$, where $\Pi_0$ is a reference permeability. We will consider two different functional forms of the permeability. A cubic permeability function 
\begin{equation} \Pi(\phi) = (1-\phi)^3 \label{eq:cubicPi} \end{equation}
is consistent with earlier studies~\citep{SchulzeWorster:1998,ChungWorster:2002} and bears close resemblance to the permeability dependence observed for sea ice over a range of solid fractions~\citep{PetrichEicken:2010}. Following  \cite{Katz:2008tg} we also consider a selection of simulations in~\S\ref{sec:permresults} using a modified Carman-Kozeny permeability
\begin{equation} \Pi(\phi) = \frac{1}{\epsilon}\left[\frac{1}{\epsilon}+\frac{\phi^2}{(1-\phi)^3}\right]^{-1}. \label{eq:CKPi} \end{equation}
This approximates the usual Carman-Kozeny form for larger $\phi$ and the factor $\epsilon=12\Pi_0/d^2$ accounts for sidewall drag in narrow Hele-Shaw cells of width $d$ and removes a singularity as $\phi\rightarrow0$. There are six dimensionless parameters governing the system, 
\begin{align} \Rm&=\frac{g\beta(C_0-C_E)\Pi_0}{\nu V},  & \St&=\frac{L}{c_p\Gamma(C_0-C_E)}, & \Cc&=\frac{C_S-C_0}{C_0-C_E}, \label{eq:params1} \\
 \Da &= \frac{\Pi_0 V^2}{\kappa^2}, & \theta_{\infty}&=\frac{T_{\infty}-T_L(C_{o})}{T_E-T_L(C_{o})}, & \lambda&=\frac{V l}{2\kappa}, \label{eq:params2} \end{align}
where the mushy layer Rayleigh number $\Rm$ describes the ratio of buoyancy to dissipation, and the Darcy number $\Da$ characterizes the mushy layer permeability. The Stefan number $\St$, concentration ratio $\Cc$, in which $C_S$ is the solid concentration, and scaled temperature $\theta_{\infty}$ characterize the imposed thermodynamic conditions. The dimensionless chimney half-spacing is $\lambda$.
Taking the curl of~(\ref{eq:Darcyu}\textit{a}) yields the vorticity equation
\begin{equation} \nabla^2 \psi = -\Rm\, \Pi \partiald{\theta}{x} -\left(\frac{\nabla \psi \cdot \nabla \Pi}{\Pi}\right)  \label{eq:vorticity} \end{equation}
where the streamfunction $\psi$ is defined by $\uvec=(-\partial\psi/\partial z,\partial \psi/\partial x)$, thereby satisfying~(\ref{eq:Darcyu}\textit{b}).
 
Flow features in the chimneys and overlying liquid can occur on smaller length scales than features in the mushy region, and so a considerable increase in computational effort would be required to fully resolve the liquid regions numerically~\cite[e.g.][]{ChungWorster:2002}. In this study, we focus on dynamics dominated by the mushy region, and utilize asymptotic and perturbation approximations  to describe the influence of the overlying liquid and chimney flow via boundary conditions for the mushy region. The boundary conditions at the mush-liquid interface $z=h$ describe coupling to the overlying liquid. If solute diffusion is neglected, the liquid region has uniform concentration $C_0$ outside of the compositional plumes exiting the mushy layer~\cite[][]{SchulzeWorster:1998}, and we assume there is constant pressure at the mush-liquid interface~\cite[][]{EmmsFowler:1994}. The mush-liquid interface advances until it eliminates constitutional supercooling, so that the free-boundary position $z=h(x,t)$ is determined from the condition of marginal equilibrium $T=T_L(C_0)$. Combining continuity of temperature and solute concentration, the above assumptions yield 
\begin{subeqnarray} \gdef\thesubequation{\theequation \textit{a,b,c}} \theta=0, \qquad \phi=0, \qquad \nvec \cdot \nabla \psi=0, \qquad \mbox{at} \qquad z=h, \label{eq:upperbcs}  \end{subeqnarray}
 \returnthesubequation
%\begin{equation} \theta=0, \qquad \phi=0, \qquad \nvec \cdot \nabla \psi=0, \qquad \mbox{at} \qquad z=h, \label{eq:upperbcs} \end{equation}
where  Darcy's law~(\ref{eq:Darcyu}\textit{a}) has been used to write the pressure condition in terms of $\psi$. The final boundary condition at the mush-liquid interface comes from continuity of normal heat fluxes  $\left. \nvec \cdot \nabla T\right|_{-}^{+}=0$.  Balancing advection and diffusion of heat across a thermal boundary layer where isotherms have curvature $\nabla \cdot \nvec$ yields
\begin{equation}\nvec \cdot \nabla \theta=\theta_{\infty} \left[  \nabla \cdot \nvec-\left(\uvec - \mathbf{k} \right) \cdot \nvec  \right], \qquad \qquad (z=h).  \label{eq:lidupdate} \end{equation}
\cite[see][for a full derivation.]{ChungWorster:2002}

We describe the flow in the chimney by combining features of previous analyses~\citep{SchulzeWorster:1998,ChungWorster:2002,Wellsetal:2010}, taking advantage of the observation that chimneys are narrow with $a\ll h$ to derive representative singular-interface conditions that are applied at $x=0$. Note that the singular-interface approximation can formally be justified via a stretched co-ordinate transform of the form $\tilde{x}=\lambda(x-a)/(\lambda-a)$ in the horizontal, and gives rise to corrections of $O(a)\ll 1$ in the governing partial differential equations. Neglecting these small corrections of $O(a)$ recovers the structure of the original partial differential equations with $\tilde{x}$ replacing $x$, but with the mushy layer occupying $0<\tilde{x}<\lambda$.  However, in what follows we will drop the $\tilde{x}$-notation and proceed with notation where the mushy region occupies $0<x<\lambda$ with the chimney described by boundary conditions applied at a singular interface at $x=0$. The chimney width $a(z,t)$ then acts as a parameter in the boundary conditions along the singular interface at $x=0$. Fluid flow in the chimney is described by lubrication theory~\citep{ChungWorster:2002}, yielding 
\begin{equation}\psi=\left[\frac{a^3}{3 \Da\Pi}+a\right]\frac{\partial \psi}{\partial x} + \frac{3}{20}\frac{\Rm}{\Da} a^3 (\theta+1), \qquad (x=0), \label{eq:chimneypsi} \end{equation}
where a quadratic Polhausen approximation has been used to determine the pre-factor in the buoyancy force~\citep{SchulzeWorster:1998}. The mass-flux boundary condition~\eqref{eq:chimneypsi} is applied to the vorticity equation~\eqref{eq:vorticity}. The term $a\partial\psi/\partial x$ in~\eqref{eq:chimneypsi} results from assuming that, in the limit of vanishing solid fraction, the velocity is continuous at the mush-liquid interface that forms the chimney wall~\citep{ChungWorster:2002}. \cite{LeBarsWorster:2006} demonstrated that an improved porous medium boundary condition imposes continuity of $\mathbf{u}$ at a small distance $\delta$ of order the pore scale inside the porous medium. However, for simplicity we here retain the boundary condition~\eqref{eq:chimneypsi} which can be viewed as the leading term in a Taylor expansion for $\delta \ll 1$.

The heat conducted into the chimney  wall balances the heat flux advected along the chimney, yielding 
\begin{equation} \partiald{\theta}{x}=\psi \partiald{\theta}{z}, \qquad \qquad (x=0).\label{eq:chimneyheatflux} \end{equation}
The heat flux boundary condition~\eqref{eq:chimneyheatflux} is applied to the heat equation~\eqref{eq:heat}. As demonstrated in appendix~\ref{app:lubrication}, this boundary condition relies on the approximation $a \psi/h\ll1$. This approximation was previously justified in the asymptotic limit $\Rm^{4/3}\Da\ll 1$ by~\cite{SchulzeWorster:1998}. The simulations presented here extend over a wider range of parameter space, and so we proceed under the \textit{ansatz} $a \psi/h\ll1$ and treat \eqref{eq:chimneyheatflux} as a leading order perturbation expansion. Our subsequent simulations reveal $a/h<0.06$ and $\psi<1$, so this is a self-consistent assumption.

The chimney width $a(z,t)$ is determined from marginal equilibrium, which for a free boundary with net outflow, yields
 \begin{equation} \partiald{\theta}{t}-\partiald{\theta}{z}+\uvec \cdot \nabla \theta=0, \qquad \qquad (x=0),\label{eq:chimneymargeq} \end{equation}
 \cite[][]{SchulzeWorster:2005}. The chimney width $a(z,t)$ adjusts in order to satisfy the condition (2.14). This is the time-dependent generalization of the marginal equilibrium condition of~\cite{ChungWorster:2002}.

The remaining boundary conditions on the lower and right-hand boundaries of the mushy region are
\begin{align} \theta&=-1, & \qquad \psi&=0, \qquad &\mbox{at} \qquad z&=0, \label{eq:basebcs} \\
\partiald{\theta}{x}&=0, &\qquad \psi&=0, \qquad &\mbox{at} \qquad x&=\lambda,\label{eq:symmetrybcs} \end{align} 
corresponding to no mass flux across the lower heat exchanger which is fixed at the eutectic temperature, and symmetry conditions at $x=\lambda$. 

\revadd{We pause here to consider some restrictions regarding the current modelling framework using ideal mushy layer theory. The use of Darcy's law to describe porous medium flow assumes a low Reynolds number $\mathrm{Re}_p=\hat{U}\hat{\delta}/\nu=O(1)$ within the pore space, where $\hat{U}$ is an approximate dimensional velocity scale and $\hat{\delta}$ an approximate lengthscale of the pores. In addition, the continuum conservation equations are justified by volume averaging across solid and liquid phases, and assume that all macroscopic flow features within the mushy layer occur on lengthscales larger than the pore scale. We return to these conditions in \S\ref{sec:optimalnumerics}.}

\subsection{Numerical methods\label{sec:numerics}}
 
 We solved the system~\eqref{eq:heat}--\eqref{eq:symmetrybcs} numerically using second-order finite differences for the time-dependent problem, complemented by an arc-length continuation method that traces both stable and unstable solution branches in order to verify the bifurcation structure. A brief summary follows, with a more detailed elaboration in Appendix~\ref{app:numerics}. 
 
 Semi-implicit Crank-Nicholson time-stepping was used to update the heat and concentration equations~\eqref{eq:heat} and~\eqref{eq:salt}, while the vorticity equation~\eqref{eq:vorticity} yields a nonlinear Poisson equation for $\psi$. After time discretization, the spatial derivatives of~\eqref{eq:heat} and~\eqref{eq:vorticity} yield elliptic systems that were solved using multigrid iteration~\cite[][]{Adams:1989,BriggsHensonMcCormick:2000}. The free boundaries $a(z,t)$ and $h(x,t)$ were updated via relaxation, with $a(z,t)$ updated at each value of $z$ to reduce the error in~\eqref{eq:chimneymargeq}. A corresponding free-boundary problem for $h(x,t)$ using~(\ref{eq:upperbcs}\textit{a}) yielded an unstable scheme when used in conjunction with the boundary layer approximation~\eqref{eq:lidupdate}. Hence, following~\cite{SchulzeWorster:1998}, a one-parameter shape
 \begin{equation} h=h_o-\frac{\psi_c \left[ 1-\mathrm{cosh}\, \mu \left(\lambda-x\right)\right]}{\mu \mathrm{sinh}\, \mu \lambda}, \qquad \psi_c=\left. \psi \right|_{x=0,z=h} \label{eq:hshape} \end{equation}
 is enforced, which has a thermal boundary layer in order to remove a temperature singularity at the chimney top. We use a thermal-boundary-layer width $1/\mu=\Rm^{-2/3}$ following the scaling analysis of~\cite{SchulzeWorster:1998}. Note that~\eqref{eq:hshape} is an approximation to $h(x,t)$ and does not provide an exact solution of the free boundary problem. We choose to apply the boundary conditions so that the condition of marginal equilibrium~(\ref{eq:upperbcs}\textit{a}) is enforced exactly at all times by setting $\theta=0$ at $z=h$ as a boundary condition on \eqref{eq:heat}, and then $h_0$ is updated to minimize the least-square residual in the heat flux condition~\eqref{eq:lidupdate}. In this manner, errors from the two approximations~\eqref{eq:lidupdate} and~\eqref{eq:hshape} are incorporated in one condition. The time-dependent initial value problem was integrated to a steady state with initial conditions  given either by an analytic solution for solidification with no fluid flow~\cite[e.g.][]{Worster:1991} or by continuation from a previous solution with different parameters. In addition to other flow quantities, we diagnose the solute flux from the mushy layer into the fluid per unit width. The combination of conservation of solute with continuity of $C$ at a freezing interface with outflow results in $\phi=0$ at the chimney wall~\citep{SchulzeWorster:2005}, and further noting that $\theta=\phi=0$ at $z=h$, the resulting solute flux reduces to a line integral over the chimney boundary, with absolute magnitude
 \begin{equation} F_s= \left| \frac{1}{\lambda}\int_{\{x=a(z)\}} \theta \left(\mathbf{u}-\mathbf{k}\right)\cdot \mathbf{n}\,dl. \right|\label{eq:Fsdef} \end{equation}
In steady state, integrating~\eqref{eq:salt} over the area of the mushy region, using the divergence theorem and applying the boundary conditions~\eqref{eq:upperbcs}, \eqref{eq:basebcs} and~\eqref{eq:symmetrybcs}  yields the alternative expression
 \begin{equation}F_s=\frac{1}{\lambda}\int_{0}^{\lambda} \left[(1-\phi)\theta+\Cc\phi \right]_{z=0}\,dx, \label{eq:Fsdef2} \end{equation}
 which is adopted here for numerical convenience.
 
 To complement the time-dependent solutions, an arc-length continuation scheme was applied to trace both stable and unstable steady states, and verify the bifurcation structure~\citep{Keller:1977}. To reduce the computational cost associated with the size of the Jacobian matrices required for arc-length continuation, we project the finite-difference solutions of~\eqref{eq:heat}--\eqref{eq:symmetrybcs} onto a basis of Chebyshev polynomials yielding a low-order pseudo-spectral representation for use in the predictor-corrector scheme. The resulting updated Chebyshev projections were then used as initial conditions in an iterative time-independent version of the finite difference code for post processing (see  Appendix~\ref{app:numerics} for further details).
 
 \section{Flow dependence on chimney spacing \label{sec:lambdaresults}}
 
\begin{figure}
 \centering
 \psfrag{lu}{$(\lambda_u,0)$}
   \psfrag{lcFc}{$(\lambda_c,F_c)$}
      \psfrag{loFo}{$(\lambda_o,F_o)$}
    \psfrag{lambda}{$\lambda=l V/2 \kappa$}
    \psfrag{Fs}{$F_s$}
                                        \includegraphics[height=5.5cm]{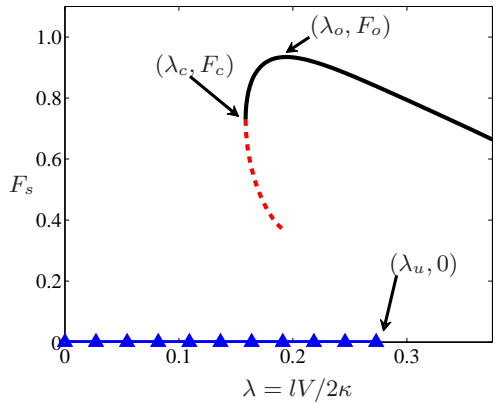}
\caption{Calculated steady-state solute flux per unit width $F_s$ as a function of chimney half-spacing $\lambda$. The steady states show hysteresis, with a stable upper branch of convection with chimneys for $\lambda>\lambda_c$ (black curve) and a  lower branch of no flow for $\lambda<\lambda_u$ (blue line with triangle symbols) separated by an unstable solution branch (red dashed curve). A fold bifurcation is observed at $(\lambda_c,F_c)$, and the optimal flux $F_o$ is achieved with chimney half-spacing $\lambda_o$. These fluxes are calculated for $\Rm=95$, $\Cc=5$, $\St=5$, $\theta_{\infty}=0.4$ and $\Da=5 \times 10^{-3}$.}
\label{fig:fluxvslambda}
\end{figure}

 We begin by considering the evolution of flow dynamics as the chimney spacing is varied, illustrating key features described by~\cite{Wellsetal:2010} to provide context for the later discussion. To demonstrate the system dynamics, figure~\ref{fig:fluxvslambda} shows a characteristic example of the variation of the dimensionless solute flux per unit width $F_s$ with chimney half-spacing $\lambda$.
The system has two stable states showing hysteresis: a lower branch of no flow for $\lambda<\lambda_u$ (solid line with triangle symbols) and an upper branch of chimney 
 convection for $\lambda>\lambda_c$ (solid black curve). The arc-length continuation method also reveals an intermediate unstable branch of chimney convection (dashed curve), with the two states of chimney convection becoming extinct via a bifurcation at $(\lambda_c,F_c)$ as $\lambda$ is reduced. For large values of $\Rm$ this corresponds to a saddle-node bifurcation. For much smaller $\Rm$, close to the linear critical value of Rayleigh number, the arc-length continuation suggests a singular Jacobian matrix at the turning point consistent with a more complex bifurcation structure involving flow without steady-state chimneys~\cite[see for example the bifurcation to a state of convection without a chimney shown in figure 2b of][]{Wellsetal:2010}. However, we emphasize that, for all simulations considered, the solute flux shows the same qualitative turning point structure as illustrated in figure~\ref{fig:fluxvslambda}, with steady states of chimney convection ceasing to exist for $\lambda<\lambda_c$. Failure of convergence of the arc-length scheme prevents continuation of the unstable branch to larger $\lambda$. 
 The simulations considered here show that this form of hysteresis behaviour identified by~\cite{Wellsetal:2010} persists for a wide range of concentration ratios $\Cc$ and Rayleigh numbers $\Rm$. 
 
\begin{figure}
 \psfrag{x}{$x$}
 \psfrag{z}{\hspace{-0.1cm} $z$}
 \psfrag{Phi}{\hspace{0.05cm} $\phi$}
 \psfrag{Th}{\hspace{0.05cm} $\theta$}
 \psfrag{eps}{$a$}
 \psfrag{a}{(\textit{a})}
  \psfrag{b}{(\textit{b})}
   \psfrag{c}{(\textit{c})}
             \includegraphics[height=14cm,angle=270]{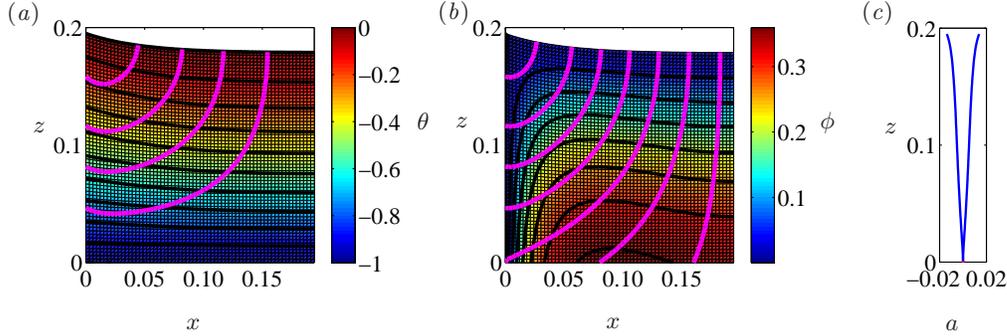}
\caption{Steady-state mushy layer profiles with an optimal chimney half-spacing $\lambda_o=0.194$, and all other parameters identical to figure~\ref{fig:fluxvslambda}.  In panel (\textit{a}) the colour scale shows dimensionless temperature $\theta(x,z)$ and solid black curves show isotherms, whilst in panel (\textit{b}) the colour scale shows solid fraction $\phi(x,z)$, with black solid contours of constant $\phi$. Magenta curves show streamlines of Darcy velocity $\mathbf{u}$ in panel (\textit{a}) and net fluid flux $\mathbf{q}=\mathbf{u}-\mathbf{k}$ in panel (\textit{b}), with inflow at the upper boundary of the mush, and outflow into the chimney at the left boundary. Panel (\textit{c}) illustrates variation of the chimney width $a(z)$ with height.}
\label{fig:profiles}
\end{figure}

 The solute fluxes illustrated in figure~\ref{fig:fluxvslambda} have an optimal value at $\lambda=\lambda_o$. Figure~\ref{fig:profiles} shows the corresponding profiles of temperature (colour scale in panel \textit{a}), solid fraction (colour scale in panel \textit{b}) and chimney width (panel \textit{c}) in this optimal state, along with streamlines of Darcy velocity (magenta curves in panel \textit{a}) and net fluid flux $\mathbf{q}=\mathbf{u}-\mathbf{k}$ relative to the heat exchangers (magenta curves in panel \textit{b}), where $\mathbf{k}$ is a unit vector in the $z$-direction.
 These profiles exhibit the same flow structure observed by \cite{Wellsetal:2010}, with inflow at the top boundary, circulation in an order-one aspect ratio convective cell, and then outflow into the chimney at the left side of the domain. At the large Rayleigh number of this simulation we observe a tapered chimney shape which is narrow at the base, differing slightly from the steep-sided chimneys observed at low Rayleigh number~\cite[see][and also confirmed using our current method]{ChungWorster:2002}. In \S\ref{sec:optimalresults} we determine  a transition in scaling regimes for the chimney \revadd{between dominance of plug flow and Poiseuille} flow, which may underlie this difference in shape. 
 
\begin{figure} 
 \psfrag{a}{(\textit{a})}
  \psfrag{b}{(\textit{b})}
      \psfrag{lambda}{$\lambda$}
            \psfrag{lambdac}{$\lambda_c$}
 %      \psfrag{ilambda}{$1/\lambda_c$}
             \psfrag{iRm}{$1/\Rm$}
       \psfrag{Rm}{$\Rm$}
       \psfrag{Cceq50}{$\Cc=50$}
        \psfrag{Cceq25}{$\Cc=25$}
        \psfrag{Cceq10}{$\Cc=10$}
        \psfrag{Cceq5}{$\Cc=5$}
        \psfrag{Cceq4}{$\Cc=4$}
        \psfrag{Cceq3}{$\Cc=3$}
        \psfrag{Cceq2}{$\Cc=2$}   
        \psfrag{CHIMNEY}{{\footnotesize CHIMNEY}}
        \psfrag{CONVECTION}{{\footnotesize CONVECTION}} 
        \psfrag{NOSTEADYCHIMNEY}{\footnotesize NO STEADY CHIMNEYS}   
                                                    \includegraphics[width=13.75cm,angle=0,trim= 5mm 2mm 4mm 2mm, clip=true]{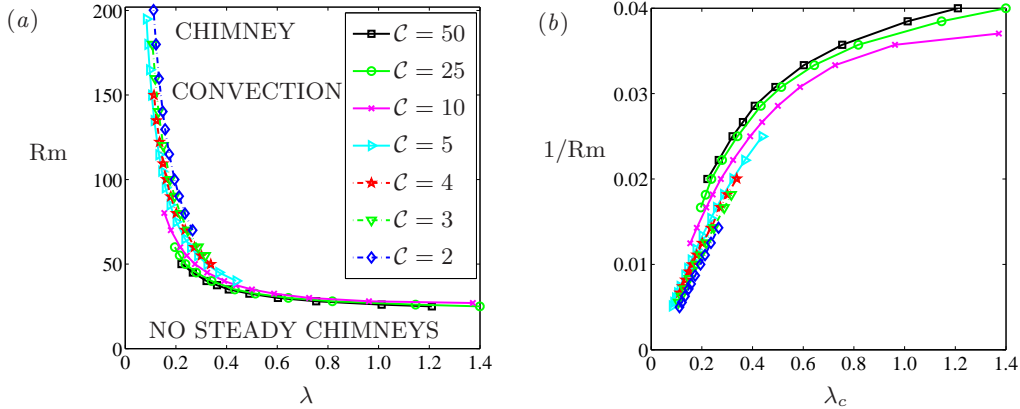}
\caption{(\textit{a}) Stability boundaries $\lambda_c(\Rm)$ for a wide variety of concentration ratios $\Cc$ as given in the legend. For a given value of $\Cc$, stable states of chimney-convection exist everywhere above the plotted section of curve (i.e. $\lambda>\lambda_c(\Rm)$) in the $(\lambda,\Rm)$ plane. (\textit{b}) Scaling of stability boundaries illustrated by re-plotting $1/\Rm$ versus $\lambda_c$ for all data in panel (\textit{a}). Line styles and symbols follow the legend in panel (\textit{a}). In both plots, all other parameters $\St=5$, $\theta_{\infty}=0.4$ and $\Da=5 \times 10^{-3}$ are held fixed.}
 \label{fig:stabplanes}
\end{figure}

 We now examine in detail the flow dynamics as the Rayleigh number and concentration ratio are varied. For fixed Rayleigh number, the bifurcation at $(\lambda_c,F_c)$ defines a stability boundary for the existence of the observed chimney-convection state in cells of confined width. Figure~\ref{fig:stabplanes} traces the stability boundaries $\lambda_c(\Rm)$ as a function of Rayleigh number for a wide range of concentration ratios with $\Cc\geq2$. 
Chimney convection is stable above each of the curve sections plotted in the $(\lambda,\Rm)$ plane in figure~\ref{fig:stabplanes}\textit{a}. All concentration ratios used are consistent with the same underlying pattern, with chimney convection suppressed for small Rayleigh numbers or small chimney spacings. This behaviour is consistent with the experiments of~\cite{Zhongetal:2012}. Note that convergence issues in the arc-length continuation method prevented us from extending the range of these curves:  the failure of convergence of the corrector scheme is responsible for failure at larger Rayleigh numbers, and for the smaller Rayleigh numbers, due to the existence of a nearby oscillatory state, Êthere was a failure of convergence of the finite-difference 
code used to generate an initial guess for the tangent vector. Figure~\ref{fig:stabplanes}\textit{b} examines the scaling of the stability boundary by plotting pairs of values $(\lambda_c,1/\Rm)$. For $\lambda_c\ll1$ the variation of $1/\Rm$  with $\lambda_c$  is approximately linear for each curve, with the $\lambda_c$-dependence weakening for larger $\lambda_c$. Figure~\ref{fig:stabplanes}\textit{b}  also demonstrates that the critical Rayleigh number for convection $\Rm_c(\lambda_c)$ decreases slightly as the concentration ratio increases, so that chimney convection is stable over a wider range of $(\lambda,\Rm)$ for larger $\Cc$.  The leading order linear variation for $\lambda_c\ll1$ is consistent with a scaling
\begin{equation} \lambda_c(\Rm) \sim \mathrm{R}_c/\Rm, \label{eq:lambdacscale} \end{equation}
for $\mathrm{R}_c$ constant, which we can reinterpret in terms of an approximate revised stability condition
\begin{equation}  \Rm_{E} \equiv \Rm \,\lambda_c \approx \mathrm{R}_c,  \label{eq:criticalRm} \end{equation}
in terms of an effective Rayleigh number
\begin{equation} \Rm_{E} \equiv \frac{g\beta(C_0-C_E)\Pi_0}{\nu \kappa}\left(\frac{l}{2}\right) , \label{eq:modifiedRayleigh} \end{equation}
formed using the permeability $\Pi_0$ and the horizontal width of a convection cell $l/2$. This modified Rayleigh number can be rationalized by noting that the mush-liquid interface is a free boundary and so the dimensional thickness $\hat{h}$ is free to evolve in response to the flow. Hence, when determining a critical porous-medium Rayleigh number $g\beta(C_0-C_E)\Pi_0 \hat{h}/\nu\kappa$ for convection, $\hat{h}$ can scale either with the vertical thermal diffusion length scale $\kappa/V$, with the convection-cell width $l/2$, or else some combination of the two. For large Rayleigh number the mushy-layer structure is dominated by the convective flow and thus we expect $\hat{h}$ to be independent of the thermal diffusion scale. This leads to $\hat{h}\propto l/2$ for $\Rm\gg1$, yielding the effective Rayleigh number~\eqref{eq:modifiedRayleigh}. Hence, for strong convection in cells of confined horizontal width, it is the horizontal length scale that provides the dominant restriction when determining allowed modes of convection. This effect is associated with the competition between convection cells for material to transport out of the mushy layer.  We expect and observe the scaling~\eqref{eq:criticalRm}  to break down when $\lambda=lV/2\kappa=O(1)$, when the vertical thermal length scale becomes comparable to the chimney spacing and plays a more significant role in controlling the stability of convecting states. 

 \section{Scaling of the optimal state for large Rayleigh number \label{sec:optimalresults}}

We now investigate how the states with optimal potential energy fluxes evolve as the vigour of convection increases for large Rayleigh number. We begin by identifying relevant scaling regimes in the governing equations to provide context for later discussion of the numerical results.

\subsection{Scaling analysis \label{sec:scaling}}

We investigate the scaling of the optimal state for strong convection in the asymptotic limit $\Rm\gg 1$, with $\Cc>1$ and $\St>1$ relevant to our numerical simulations. In particular, consistent with our hypothesis that the chimney spacing adjusts to provide an optimal solute flux, we derive scalings where $\lambda$ varies with $\Rm$ (other scalings are possible if the chimney spacing is specified \textit{a priori} and introduces an additional length scale into the problem). Noting that $\theta=O(1)$ as a result of the boundary conditions and $0<\phi<1$ by definition, we determine the possible self-consistent asymptotic scalings of~\eqref{eq:heat}--\eqref{eq:symmetrybcs} that take the form 
\begin{equation} \psi =\Rm^{b} \Psi, \quad x=\Rm^{-c} X, \quad z=\Rm^{-d} Z, \quad \lambda=\Rm^{-c} \bar{\lambda}, \quad h=\Rm^{-d} H, \quad \label{eq:ansatz} \end{equation} 
capturing convection that penetrates the full depth of the mushy layer. We shall demonstrate below that such scalings will only be valid with the further condition $\Rm/\Cc=O(1)$, in order to maintain a solid fraction with $0<\phi<1$. With the expectation that $X$, $Z$, $\bar{\lambda}$, $H$ and $\Psi$ will be of order one, the exponents $b$, $c$ and $d$ can be determined as follows.  The balance of conduction of heat into the chimney with heat advected along the chimney in~\eqref{eq:chimneyheatflux} requires $c=d+b$, yielding
\begin{equation} \partiald{\theta}{X}=\Psi \partiald{\theta}{Z}, \qquad\qquad (X=0).\label{eq:rescalechimneyheatbc} \end{equation}
At the mush--liquid interface, the balance of heat advected into the boundary layer with 
heat conducted into the mush requires $d=c+b$, so that we must have $c=d$, $b=0$, and hence~\eqref{eq:lidupdate} reduces to
\begin{equation}\nvec \cdot \hat{\nabla} \theta=\theta_{\infty} \left( \hat{\nabla} \cdot \nvec-\mathbf{U}\cdot \nvec   \right) +O\left(\frac{1}{\Rm^c}\right), \qquad \qquad (Z=H),  \label{eq:rescalelidupdate} \end{equation}
where $\hat{\nabla}=(\partial/\partial X, \partial/\partial Z)$ and $\mathbf{U}=(-\partial \Psi/\partial Z,\partial \Psi/\partial X)$. Next, noting that $\Pi=O(1)$ near the mush--liquid interface, a balance of baroclinic torque with viscous dissipation of vorticity in~\eqref{eq:vorticity} requires $c=1$, yielding
 \begin{equation}
 \hat{\nabla}^2 \Psi = - \Pi \partiald{\theta}{X} -\left(\frac{\hat{\nabla} \Psi \cdot \hat{\nabla} \Pi}{\Pi}\right).  \label{eq:rescalevorticity} \end{equation}
These conditions result in the asymptotic scalings
\begin{equation} \lambda =O\left(\frac{1}{\Rm}\right), \quad h =O\left(\frac{1}{\Rm}\right), \quad \psi =O(1), \qquad \mbox{for} \quad \Rm \gg 1, \label{eq:hlampsiscale} \end{equation}
from which we expect that $\mathbf{u}=(-\partial \psi/\partial z,\partial \psi/\partial x)=O(\Rm)$ and from~\eqref{eq:Fsdef} the solute flux will have the scaling
\begin{equation} F_s =O(\Rm), \label{eq:fluxscale} \end{equation}
consistent with the observation of~\cite{Wellsetal:2010}.

The rescaled versions of the heat equation~\eqref{eq:heat} and solute equation~\eqref{eq:salt} become
 \begin{align}
-\frac{1}{\Rm}\partiald{\theta}{Z}+\mathbf{U} \cdot \hat{\nabla} \theta&= \hat{\nabla}^2 \theta -\frac{\St}{\Rm}\partiald{\phi}{Z},  \qquad \mbox{and} \quad
\label{eq:rescaleheat} \\
\partiald{}{Z}\left[ \phi+\frac{(1-\phi)\theta}{\Cc} \right] &= \frac{\Rm}{\Cc}\mathbf{U}\cdot \hat{\nabla} \theta. \label{eq:rescalesalt}
\end{align}
If $\mathbf{U}\cdot \hat{\nabla} \theta\neq 0$ somewhere in the domain, then for $\Rm\gg1$ \eqref{eq:rescalesalt} can only be satisfied in a self-consistent fashion if $\Rm/\Cc=O(1)$, so that $\Cc\gg 1$. Hence \eqref{eq:rescalesalt} predicts a solid fraction scale
\begin{equation} \phi =O\left(\frac{\Rm}{\Cc}\right). \label{eq:phiscale} \end{equation}
This balance breaks down for $\Rm \gg \Cc$. In particular, we expect that as $\phi$ approaches $1$, the permeability $\Pi\rightarrow 0$ and reduces the strength of flow given by the rescaled vorticity equation~\eqref{eq:rescalevorticity}. This form of scaling analysis cannot lead to scaling laws suitable for $\Rm\gg\Cc$, and it is possible that the variation in permeability may lead to convective flow that does not penetrate the full depth of the mushy layer. The scaling~\eqref{eq:phiscale} can be used to show that~\eqref{eq:rescaleheat} yields a self-consistent balance provided $\St/\Cc=O(1)$.

The boundary conditions~\eqref{eq:basebcs}-\eqref{eq:symmetrybcs} at $z=0$ and $x=\lambda$ retain the same form after rescaling, and at the chimney wall~\eqref{eq:chimneymargeq} yields 
\begin{equation} \mathbf{U}\cdot \hat{\nabla} \theta= O\left(\frac{1}{\Rm}\right), \qquad\qquad (X=0).\label{eq:rescalechimneybcs2} \end{equation}
and the mass flux condition yields
\begin{equation}\Psi=\frac{\Rm}{\Da} a^3\left[\frac{3}{20} (\theta+1)+\frac{1}{3\Pi}\frac{\partial \Psi}{\partial X}\right]+a\Rm\,\frac{\partial \Psi}{\partial X}, \qquad\qquad (X=0). \label{eq:rescalechimneypsi} \end{equation}
Note that there is upflow with $\partial \Psi/\partial X>0$ in the neighbourhood of the chimney, and hence~\eqref{eq:rescalechimneypsi} provides two possible consistent scalings for the chimney width valid for two different ranges of $\Rm^2\Da$. If $\Rm^2\Da\ll1$, then 
\begin{equation} a=O\left( \frac{\Da^{1/3}}{\Rm^{1/3}}\right), \label{eq:chimscale1} \end{equation}
and the final term in~\eqref{eq:rescalechimneypsi} is negligible. \revadd{Poiseuille flow in the chimney} is driven by buoyancy in the chimney and the pressure gradient exerted by the flow in the mush, both forces being of the same asymptotic magnitude. In this limit, the dimensional chimney width $2\hat{a}\propto  \hat{l}_f\equiv \left[\nu\kappa/g\beta(C_0-C_E)\right]^{1/3}$ has a length scale $\hat{l}_f$ relevant for convection in a purely liquid region.
The second possible scaling for $a$ occurs when $\Rm^2\Da\gg1$, and we find
\begin{equation} a=O\left(\frac{1}{\Rm}\right), \label{eq:chimscale2} \end{equation}
so that the final term dominates the right-hand side of~\eqref{eq:rescalechimneypsi} \revadd{so that there is plug flow in the chimney}. Here the chimney forms a passive conduit of enhanced permeability, adjusting to accommodate the flux supplied by convection in the mushy region. The dimensional chimney width satisfies $2\hat{a}\propto \hat{l}_p\equiv \nu\kappa/g\beta(C_0-C_E)\Pi_0$ consistent with a lengthscale $\hat{l}_p$ controlled by porous medium convection in the interior of the mushy layer. The group $\Rm^2\Da=(\hat{l}_f/\hat{l}_p)^3$ can therefore be interpreted as a measure of the relative strengths of porous medium convection compared to pure fluid convection. \revadd{We note here that whilst solutions in the limit~\eqref{eq:chimscale2} remain mathematically consistent solutions of mushy-layer theory, the manifestation of the chimney in particular physical contexts may depend on the details of the microstructure of the material in question, as discussed in more detail at the end of~\S\ref{sec:optimalnumerics}.}

It is interesting to note that if $a$ is determined consistent with the condition~\eqref{eq:rescalechimneypsi}, the asymptotic scalings of the remainder of the flow remain independent of the particular scaling of $a$.  This emphasizes the role of chimneys as passive conduits that assist the drainage of fluid driven by convection in the interior of the mushy layer. The scalings \eqref{eq:hlampsiscale}--\eqref{eq:fluxscale}, \eqref{eq:phiscale} and~\eqref{eq:chimscale1} have also been recovered independently in a recently developed simplified model of chimney convection that assumes constant permeability and a linear background temperature gradient~\citep{ReesJonesWorster:submit}, and we now compare the scalings to full numerical simulations where the assumptions of the simplified model are relaxed.

\subsection{Numerical results in the optimal state\label{sec:optimalnumerics}}

States with optimal fluxes were identified in the numerical simulations for a range of values of $\Rm$ and $\Cc$. Guided by the scaling analysis in~\S\ref{sec:scaling},  we now investigate the behaviour of the optimal state as $\Rm$ varies with all other parameters held fixed. Such variation in the vigour of convection could be achieved in directional solidification experiments by varying the growth rate $V$. \revadd{In this section we fix the parameter values $\St=5$, $\theta_{\infty}=0.4$ and $\Da=5 \times 10^{-3}$ for consistency with the previous study of~\cite{ChungWorster:2002}.}

\begin{figure}
\centering
  \psfrag{a}{(\textit{a})}
  \psfrag{b}{(\textit{b})}
  \psfrag{Cc}{$\Cc$}
  \psfrag{alp}{$\alpha$}
\psfrag{BrinefluxvsRm}{}
      \psfrag{Fs}{$F_o$}
       \psfrag{Rm}{$\Rm$}
              \psfrag{Cc75}{\scriptsize$\Cc=75$}
       \psfrag{Cc50}{\scriptsize$\Cc=50$}
        \psfrag{Cc25}{\scriptsize$\Cc=25$}
               \psfrag{Cc15}{\scriptsize$\Cc=15$}
        \psfrag{Cc10}{\scriptsize$\Cc=10$}
        \psfrag{Cc5}{\scriptsize$\Cc=5$}
        \psfrag{Cc4}{\scriptsize$\Cc=4$}
        \psfrag{Cc3}{\scriptsize$\Cc=3$}
        \psfrag{Cc2}{\scriptsize$\Cc=2$}  
             \psfrag{Rmeq40}{\scriptsize$\Rm=40$}  
                \psfrag{Rmeq50}{\scriptsize$\Rm=50$}  
                   \psfrag{Rmeq60}{\scriptsize$\Rm=60$}  
          \includegraphics[width=13.75cm,angle=0,trim= 0mm 0mm 0mm 0mm, clip=true]{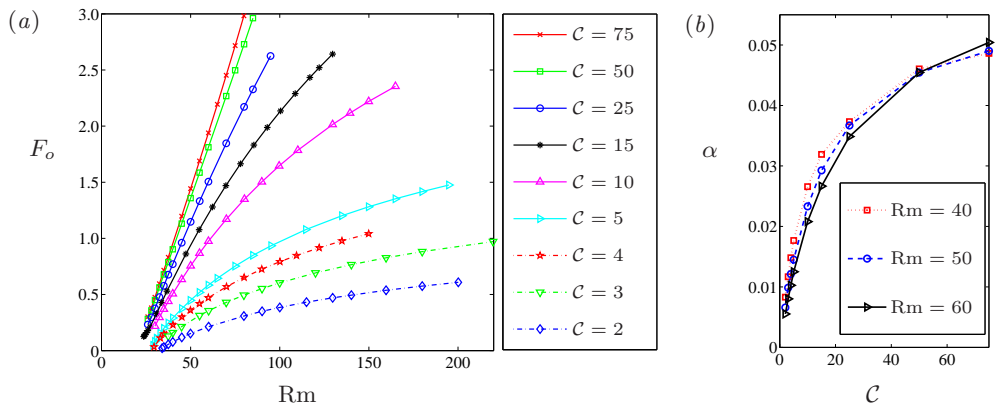}
\caption{(\textit{a}) Variation of the optimal solute flux $F_o$ with mush Rayleigh number $\Rm$, for a variety of concentration ratios~$\Cc$. Line styles and symbols are indicated in the legend. All other parameters $\St=5$, $\theta_{\infty}=0.4$ and $\Da=5 \times 10^{-3}$ are held fixed. For moderate $\Rm$ these curves can be approximated by $F_o\sim\alpha(\Rm-\Rmo)$, where $\alpha$ and $\Rmo$ are independent of $\Rm$ at leading order.  (\textit{b}) Variation of slope $\alpha(\Cc)=\partial F_o/\partial \Rm$ for three different values of $\Rm$ indicated in the legend, corresponding to the near-linear sections of the curves in panel (\textit{a}). Data points indicated by symbols are joined by linear interpolation to clarify the underlying trend. }
 \label{fig:fluxvsRm}
\end{figure}

\begin{figure} 
\centering
  \psfrag{a}{(\textit{a})}
  \psfrag{b}{(\textit{b})}
\psfrag{MaxSolidFractionvsRm}{}
      \psfrag{Phim}{$\phi_{max}$}
       \psfrag{Rm}{$\Rm$}
              \psfrag{RmoCc}{$\Rm/\Cc$}
              \psfrag{Cc75}{\scriptsize$\Cc=75$}
       \psfrag{Cc50}{\scriptsize$\Cc=50$}
        \psfrag{Cc25}{\scriptsize$\Cc=25$}
               \psfrag{Cc15}{\scriptsize$\Cc=15$}
        \psfrag{Cc10}{\scriptsize$\Cc=10$}
        \psfrag{Cc5}{\scriptsize$\Cc=5$}
        \psfrag{Cc4}{\scriptsize$\Cc=4$}
        \psfrag{Cc3}{\scriptsize$\Cc=3$}
        \psfrag{Cc2}{\scriptsize$\Cc=2$}  
                                  \includegraphics[width=13.75cm,angle=0,trim= 0mm 0mm 0mm 0mm, clip=true]{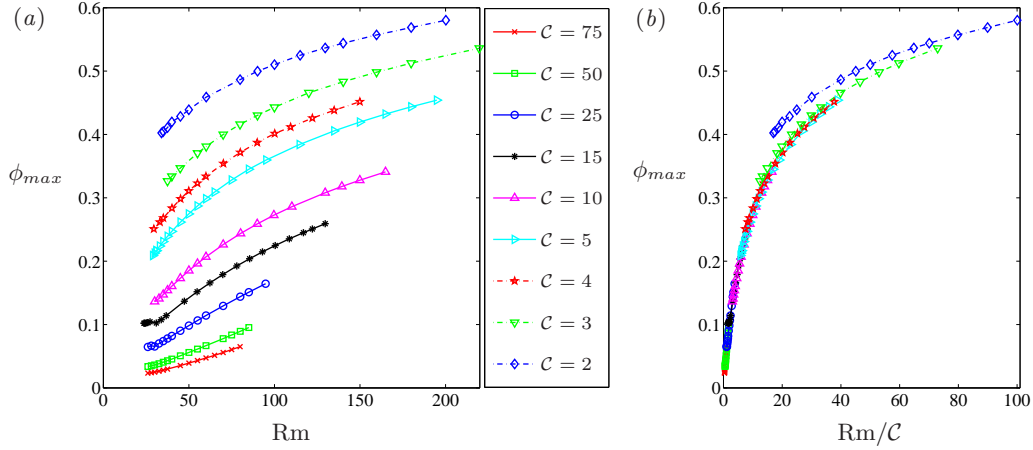}
\caption{(a) Variation of the maximum solid fraction $\phi_{max}$ in the domain with mush Rayleigh number $\Rm$, for a variety of concentration ratios~$\Cc$. (b) Collapse of the same data plotted as a function of $\Rm/\Cc$. Curves for $\Cc=3$ and $\Cc=2$ deviate slightly from the collapsed trend, which is consistent with the hypothesized breakdown of the scalings when $\Cc=O(1)$. Line styles and symbols are indicated in the legend, and are identical to figure~\ref{fig:fluxvsRm}. All other parameters $\St=5$, $\theta_{\infty}=0.4$ and $\Da=5 \times 10^{-3}$ are held fixed. Kinks arising in the curves for small $\Rm$ are associated with a discontinuous change in spatial position of the maximum value of $\phi$.}
 \label{fig:phivsRm}
\end{figure}

Figure~\ref{fig:fluxvsRm}\textit{a} shows the variation of the optimal solute flux $F_o$ with mush-Rayleigh number $\Rm$ for a wide range of concentration ratios $2\leq\Cc\leq75$.
%\begin{figure}
%\centering
%\psfrag{BrinefluxvsRm}{}
%      \psfrag{Fs}{$F_o$}
%       \psfrag{Rm}{$\Rm$}
%              \psfrag{Cc75}{\scriptsize$\Cc=75$}
%       \psfrag{Cc50}{\scriptsize$\Cc=50$}
%        \psfrag{Cc25}{\scriptsize$\Cc=25$}
%               \psfrag{Cc15}{\scriptsize$\Cc=15$}
%        \psfrag{Cc10}{\scriptsize$\Cc=10$}
%        \psfrag{Cc5}{\scriptsize$\Cc=5$}
%        \psfrag{Cc4}{\scriptsize$\Cc=4$}
%        \psfrag{Cc3}{\scriptsize$\Cc=3$}
%        \psfrag{Cc2}{\scriptsize$\Cc=2$}  
%          \includegraphics[height=6cm,angle=0,trim= 0mm 0mm 0mm 0mm, clip=true]{Images/JFMArchivalPlots/JFMPlots070312/BrineFluxvsRm}
%\caption{Variation of the optimal solute flux $F_o$ with mush Rayleigh number $\Rm$, for a variety of concentration ratios~$\Cc$. Line styles and symbols are indicated in the legend. All other parameters $\St=5$, $\theta_{\infty}=0.4$ and $\Da=5 \times 10^{-3}$ are held fixed.}
% \label{fig:fluxvsRm}
%\end{figure} 
 For the largest concentration ratios, the flux varies approximately linearly with Rayleigh number across the full plotted range. For smaller values of $\Cc$, initially the flux varies approximately linearly with Rayleigh number before the growth begins to saturate with sub-linear growth for larger values of $\Rm$. This is consistent with the scaling behaviour identified in \S\ref{sec:scaling} with $F_o \sim \alpha \Rm$ for $\Rm\gg1$ and $\Rm/\Cc=O(1)$, where the slope $\alpha$ is constant, but the scaling breaks down when $\Rm\gg\Cc$ as the solid fraction becomes large and reduces the permeability. The magnitude of the solute flux increases with $\Cc$, and in figure~\ref{fig:fluxvsRm}\textit{b} we illustrate the $\Cc$-dependence of the slope $\alpha=\partial F_o/\partial \Rm$ for three moderate values of $\Rm$ in the region of near-linear scaling where $F_o\sim\alpha(\Rm-\Rmo)$.  The increase in flux with $\Cc$ is also consistent with variation of the permeability, with the scaling~\eqref{eq:phiscale} implying smaller solid fraction and hence higher permeability for larger values of $\Cc$. Figure~\ref{fig:phivsRm}\textit{a} provides further support for these mechanisms, where we notice that the maximum solid fraction $\phi_{max}$ is larger for smaller values of $\Cc$, increases with $\Rm$, and begins to saturate for large values of $\Rm$ and small $\Cc$. 
To further investigate the scaling~\eqref{eq:phiscale}, figure~\ref{fig:phivsRm}\textit{b} illustrates the variation of the maximum solid fraction with $\Rm/\Cc$. Data for $4\leq\Cc\leq75$ collapse onto a single curve, with $\phi_{max}$ varying approximately linearly with $\Rm/\Cc$ for small $\Rm/\Cc$, consistent with the scaling~\eqref{eq:phiscale} for $\Rm/\Cc=O(1)$ and $\Rm,\Cc \gg 1$. Data for $\Cc=2$ and $\Cc=3$ follow the same qualitative trend whilst showing a slight deviation from the collapse of the remaining data. This is consistent with the hypothesized breakdown of the scalings when $\Cc=O(1)$.

\begin{figure} 
 \centering
  \psfrag{a}{(\textit{a})}
  \psfrag{b}{(\textit{b})}
   \psfrag{c}{(\textit{c})}
  \psfrag{d}{(\textit{d})}
\psfrag{InverseChimneySpacing}{}
\psfrag{InverseMushThickness}{}
\psfrag{StreamfunctionMaximum}{}
\psfrag{VerticalVelocityMaximum}{}
      \psfrag{PsiM}{$\psi_{max}$}
            \psfrag{wmax}{$\left|w_{max}\right|$}
                  \psfrag{InvL}{\quad$\displaystyle{\frac{1}{\lambda_o}}$}
                   \psfrag{InvH}{\quad$\displaystyle{\frac{1}{\bar{h}}}$}
       \psfrag{Rm}{$\Rm$}
              \psfrag{Cc75}{\scriptsize$\Cc=75$}
       \psfrag{Cc50}{\scriptsize$\Cc=50$}
        \psfrag{Cc25}{\scriptsize$\Cc=25$}
               \psfrag{Cc15}{\scriptsize$\Cc=15$}
        \psfrag{Cc10}{\scriptsize$\Cc=10$}
        \psfrag{Cc5}{\scriptsize$\Cc=5$}
        \psfrag{Cc4}{\scriptsize$\Cc=4$}
        \psfrag{Cc3}{\scriptsize$\Cc=3$}
        \psfrag{Cc2}{\scriptsize$\Cc=2$}  
                       \includegraphics[width=13.75cm,angle=0,trim= 15mm 5mm 5mm 0mm, clip=true]{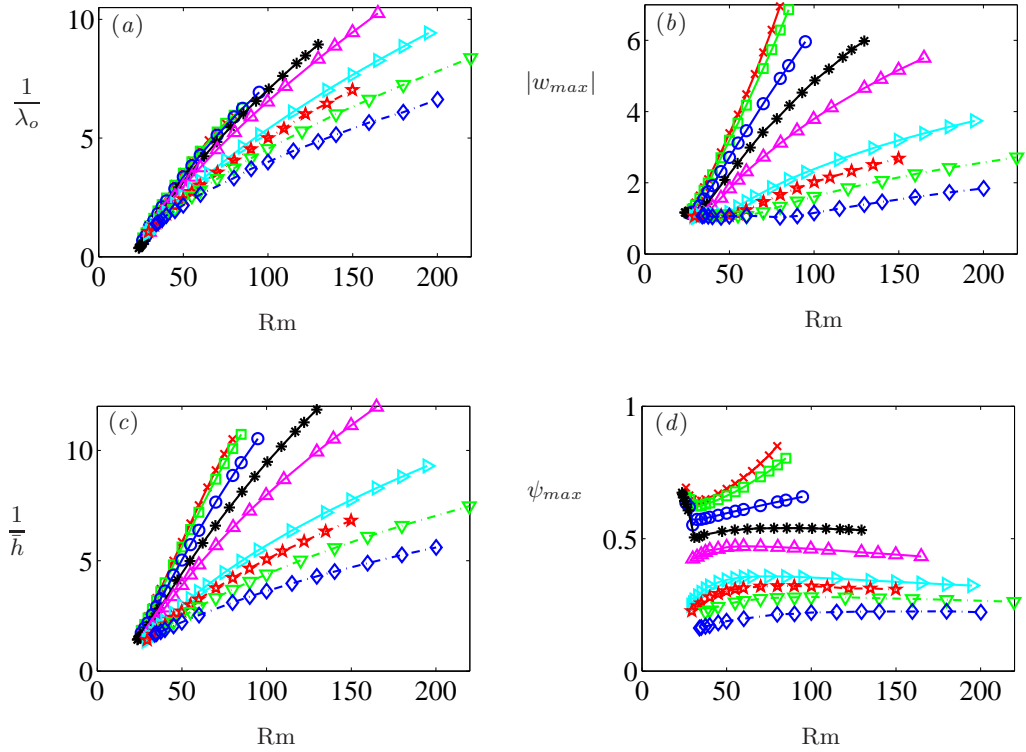}
\caption{Variation with Rayleigh number $\Rm$ of flow properties in the optimal-flux state, including (\textit{a}) inverse-chimney spacing $1/\lambda_o$, (\textit{b}) absolute maximum flow velocity $\left|w_{max}\right|$, (\textit{c}) inverse-thickness of the mushy-layer $1/\bar{h}$, (\textit{d}) maximum streamfunction value $\psi_{max}$ as a measure of fluid flux. Each curve corresponds to a different value of the concentration ratio $\Cc$, and the line styles and symbols are identical to figure~\ref{fig:fluxvsRm}, with $\Cc=2$ (blue dashed curve with $\diamond$) , $\Cc=3$ (green dashed curve with $\bigtriangledown$) , $\Cc=4$ (red dashed curve with $\star$), $\Cc=5$ (cyan solid curve with $\triangleright$), $\Cc=10$ (magenta solid curve with $\bigtriangleup$),  $\Cc=15$ (black solid curve with $\ast$),  $\Cc=25$ (blue solid curve with $\circ$), $\Cc=50$ (green solid curve with $\square$), and $\Cc=75$ (red solid curve with $\times$). All other parameters $\St=5$, $\theta_{\infty}=0.4$ and $\Da=5 \times 10^{-3}$ are held fixed. Kinks arising in the curves for small $\Rm$ are associated with a discontinuous change in spatial position of the relevant maximum value.}
 \label{fig:hlampsiwvsRm}
\end{figure}

Figure~\ref{fig:hlampsiwvsRm} demonstrates the scaling behaviour of the optimal chimney half-spacing $\lambda_o$ and corresponding mean mush thickness $\bar{h}$, maximum vertical velocity $w_{max}$ and maximum streamfunction $\psi_{max}$. 
Motivated by the scalings~\eqref{eq:hlampsiscale} we plot $1/\lambda_o$ and $1/\bar{h}$ versus $\Rm$ in panels \textit{a} and \textit{c} respectively. Both plots show an initial period of linear variation, consistent with the scalings $\lambda=O(\Rm^{-1})$ and $h=O(\Rm^{-1})$, before the previously discussed flow saturation effect leads to deviation from the linear trend at larger $\Rm$. For a given value of $\Rm$ both the optimal chimney spacing $\lambda_o$ and mean mush depth $\bar{h}$ decrease as $\Cc$ increases. Note that the scalings $\lambda\propto\Rm^{-1}$ and $h\propto\Rm^{-1}$ give rise to the $O(1)$-aspect-ratio convective cells identified by \cite{Wellsetal:2010} and~\cite{Wellsetal:2011}, and also observed in enthalpy method simulations in wide cells by~\cite{Katz:2008tg}. The absolute maximum vertical velocity $\left| w_{max}\right|$ initially increases approximately linearly with $\Rm$ (panel \textit{b}) with stronger flow for larger $\Cc$. The maximum streamfunction $\psi_{max}$ remains $O(1)$ throughout, with $\psi_{max}$ increasing slightly with $\Cc$. Note that for $25\leq \Cc\leq75$, $\psi_{max}$ increases slightly with $\Rm$ after an initial transient, whilst for $\Cc\leq15$ we observe that $\psi_{max}$ begins to decrease for large $\Rm$ consistent with saturation of the mass flux for $\Rm\gg\Cc$.  

The behaviour of $F_s$, $\phi_{max}$, $\lambda_o$, $\bar{h}$, $w_{max}$, and $\psi_{max}$ are consistent with the hypothesized scaling regime, which suggests the following picture for the development of the flow in directional solidification. As the strength of convection increases, enhanced advection of heat from the overlying liquid gives rise to a thinner mushy layer with $h\propto\Rm^{-1}$. Optimal drainage of potential energy is achieved with $O(1)$-aspect-ratio convection cells which minimize the resistance to flow, and hence the optimal chimney spacing also reduces with $\lambda\propto\Rm^{-1}$. The mass flux through each convective cell remains of similar magnitude, with $\psi_{max}=O(1)$ as $\Rm$ increases, but the smaller chimney spacing allows a higher density of convective cells so that the fluid velocity $\mathbf{u}=O(\Rm)$ and solute flux $F_s=O(\Rm)$ both increase with $\Rm$.  The velocities and fluxes are smaller for smaller $\Cc$ because the solid fraction increases as $\Cc$ decreases, and the decreased permeability provides more resistance to flow. For sufficiently strong convection with $\Rm\gg\Cc$ there is a saturation effect as the solid fraction becomes large, the permeability is reduced, and the flow and solute fluxes grow more slowly. It is important to note that this picture for directional solidification does not translate directly to the different setting of transient growth, where $h(t)$ grows over time and the observed coarsening of chimney spacing as $h(t)$ increases is consistent with the aspect ratio $\lambda/h=O(1)$~\citep{Wellsetal:2010}. 

\begin{figure} 
  \psfrag{a}{(\textit{a})}
  \psfrag{b}{(\textit{b})}
      \psfrag{IEps3}{\;\;$\displaystyle{\frac{1}{a^3}}$}
       \psfrag{IEps}{\;\;$\displaystyle{\frac{1}{a}}$}
       \psfrag{Rm}{$\Rm$}   
                                                                \includegraphics[width=13.75cm,angle=0,trim= 10mm 0mm 10mm 0mm, clip=true]{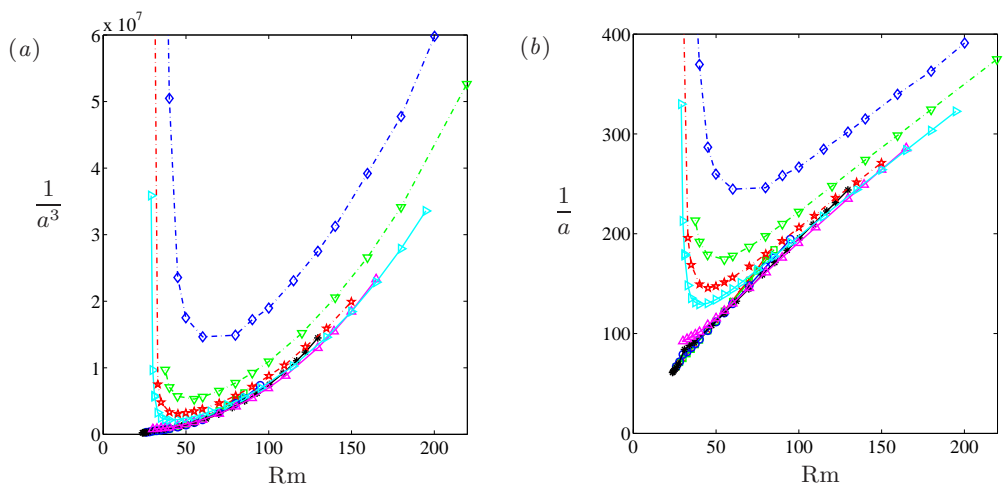}
\caption{Comparison of different scaling regimes for the chimney width $a$ in the optimal state as a function of Rayleigh number $\Rm$, with each curve representing a different concentration ratio $\Cc$ indicated by different line and symbol styles. For large $\Rm$, panel (\textit{a}) shows that $1/a^3$ increases nonlinearly with $\Rm$, whilst panel (\textit{b}) indicates that $1/a$ scales approximately linearly with $\Rm$, consistent with the scaling~\eqref{eq:chimscale2}. Curves for small $\Rm$ and $2\leq\Cc\leq5$ show a divergence of $1/a^3$ and $1/a$, because $a\rightarrow0$ and the chimneys close as the Rayleigh number approaches the critical threshold for existence of chimney convection.  The line styles and symbols are identical to figure~\ref{fig:fluxvsRm}, with $\Cc=2$ (blue dashed curve with $\diamond$) , $\Cc=3$ (green dashed curve with $\bigtriangledown$) , $\Cc=4$ (red dashed curve with $\star$), $\Cc=5$ (cyan solid curve with $\triangleright$), $\Cc=10$ (magenta solid curve with $\bigtriangleup$),  $\Cc=15$ (black solid curve with $\ast$),  $\Cc=25$ (blue solid curve with $\circ$), $\Cc=50$ (green solid curve with $\square$), and $\Cc=75$ (red solid curve with $\times$). All other parameters $\St=5$, $\theta_{\infty}=0.4$ and $\Da=5 \times 10^{-3}$ are held fixed.}
 \label{fig:avsRm}
\end{figure}

In \S\ref{sec:scaling}, we identified two different possible scalings for the chimney width given by~\eqref{eq:chimscale1} and~\eqref{eq:chimscale2}. To investigate which of these scalings is relevant we plot $a^{-3}$ versus $\Rm$ in figure~\ref{fig:avsRm}\textit{a} and $a^{-1}$ versus $\Rm$ in figure~\ref{fig:avsRm}\textit{b}.
For large values of $\Rm$, figure~\ref{fig:avsRm} shows that $a^{-3}$ increases nonlinearly with $\Rm$, whilst $a^{-1}$ shows approximately linear variation with $\Rm$. This suggests that for large $\Rm$ our numerical simulations with $\Da=5\times 10^{-3}$ correspond to the scaling $a=O(1/\Rm)$, consistent with the limit $\Rm^2\Da\gg1$. \revadd{However, one would expect the alternative scaling~\eqref{eq:chimscale1} to be achieved for smaller values of $\Da$.}   For the simulations with $2\leq\Cc\leq5$, both $1/a^3$ and $1/a$ diverge for small $\Rm$, with the chimney closing and $a\rightarrow0$ as the Rayleigh number approaches the critical threshold for existence of chimney convection. 

We conclude this section by returning to the criterion for physical applicability of ideal mushy layer theory. The scaling behaviours and numerical solutions discussed above are mathematically consistent solutions of the relevant ideal mushy layer equations, which contain no direct information about the material microstructure. However, the details of this microstructure may affect the applicability of certain aspects of the theory to particular physical settings. We focus our discussion on materials where the permeability $\Pi_o\sim \hat{\delta}^2$ scales as the square of the dimensional pore scale $\hat{\delta}$ (neglecting any pre-factor in the scaling), so that the dimensionless pore scale satisfies $\delta\sim \Da^{1/2}$. For this class of materials, we can estimate the pore-space Reynolds number as $\mathrm{Re}_p\sim\hat{U}\hat{\delta}/\nu \sim \left|w\right|\Da^{1/2}/\sigma$ where $\sigma=\nu/\kappa$ is the Prandtl number. The asymptotic scalings~\eqref{eq:hlampsiscale} yield $\mathrm{Re}_p=O\left(\Rm\,\Da^{1/2}/\sigma\right)$, so that the Darcy flow approximation breaks down when $\Rm\,\Da^{1/2}\gg \sigma$. Note that the parameter range of our numerical results leads to $\left|w\right| \Da^{1/2}<0.5$, so that the Darcy flow approximation is reasonable for any fluids of moderate to large Prandtl number. The physical details of the microstructure may also be significant for determining whether macroscopic flow features are larger than the pore scale. The chimneys provide the smallest macroscopic lengthscale in the system, with two different scalings.  The scaling~\eqref{eq:chimscale1} results in a ratio $\delta/a=O\left(\Rm^{1/3}\Da^{1/6}\right)$ so that the chimney width is wider than the pore scale in the asymptotic limit $\Rm \,\Da^{1/2}\rightarrow 0$. By contrast, the scaling~\eqref{eq:chimscale2} gives $\delta/a=O\left(\Rm\,\Da^{1/2}\right)$ so that materials with $\Pi_o\sim\hat{\delta}^2$ have a predicted chimney width smaller than the physical pore scale for sufficiently large values of $\Rm \,\Da^{1/2}$. Hence, whilst the mathematical description remains self consistent for $\Rm \,\Da^{1/2}\gg 1$, the physical manifestation of the chimney shape in the limit~\eqref{eq:chimscale2} may depend on the permeability and microstructure of the material in question.  However, we emphasize that the asymptotic flow structure in the interior of the mushy layer remains relatively insensitive to the particular shape of the chimney with the scalings~\eqref{eq:hlampsiscale}, \eqref{eq:fluxscale}, and \eqref{eq:phiscale} for the interior of the mushy layer remaining valid in both regimes~\eqref{eq:chimscale1} and~\eqref{eq:chimscale2} for the chimney.  In particular, the value of $\Da$ only enters the problem through \eqref{eq:chimneypsi}, and hence if chimney scalings are determined consistent with~\eqref{eq:chimneypsi} we expect the dominant structure of the flow in the interior of the mushy layer to be robust, and independent of both the particular chimney shape and value of $\Da$. This would be the case if one were to solve the heat equation~\eqref{eq:heat} and the vorticity  equation~\eqref{eq:vorticity} subject to boundary conditions~\eqref{eq:chimneyheatflux} and~\eqref{eq:chimneymargeq}, with $a$ adjusted to satisfy~\eqref{eq:chimneypsi}. The physics of the initial problem and our numerical methodology correspond to a solution of the heat equation~\eqref{eq:heat} and vorticity equation~\eqref{eq:vorticity} using boundary conditions~\eqref{eq:chimneyheatflux} and~\eqref{eq:chimneypsi} at $x=0$, with $a$ adjusted to satisfy the marginal equilibrium condition~\eqref{eq:chimneymargeq}. Whilst mathematically rational, we have not proved that these two different approaches are guaranteed to give the same solution. However, it is interesting to note that previous simulations that varied $\Da$ with all other parameters held fixed resulted in a mass flux through the chimney wall that was independent  of $\Da$~\citep{Zhongetal:2012}. 

\section{Dependence of flow on the permeability function \label{sec:permresults}}

\begin{figure} 
 \psfrag{a}{(\textit{a}) Cubic permeability}
  \psfrag{b}{(\textit{b}) C-K}
      \psfrag{lambda}{$\lambda$}
       \psfrag{Fs}{$F_s$}    
                                        \includegraphics[height=14cm,angle=270,trim= 0mm 0mm 0mm 0mm, clip=true]{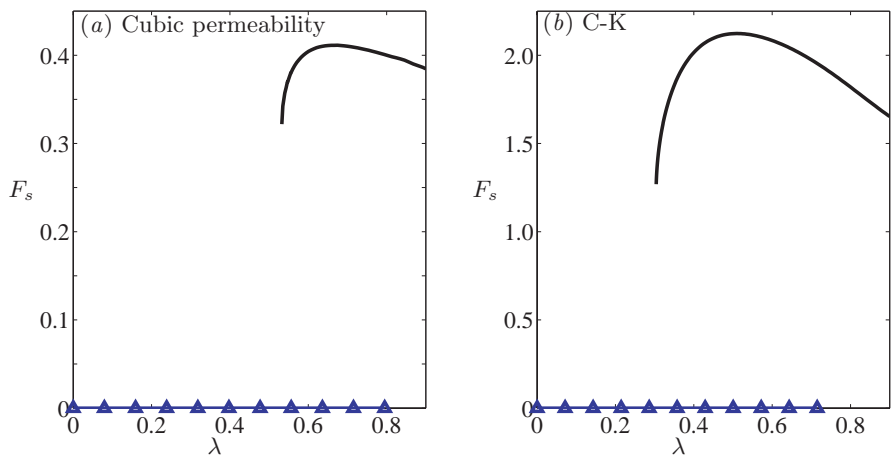}
\caption{Comparison of stable steady states of solute flux $F_s$ as a function of chimney half-spacing $\lambda$ using (\textit{a}) the cubic permeability function~\eqref{eq:cubicPi}, and (\textit{b}) the Carman-Kozeny (C-K) permeability function~\eqref{eq:CKPi} using $\epsilon=0.1$. Both permeability functions result in hysteretic behaviour, with an upper stable branch of chimney convection for $\lambda>\lambda_c$ (black curve) and a lower branch with a state of no flow being stable for $\lambda<\lambda_u$ (blue line with triangles). The parameters $\Rm=35$, $\Cc=15$, $\St=3$, $\theta_{\infty}=0.5$ and $\Da=10^{-4}$ are held fixed in both plots.}
 \label{fig:FluxPermComp}
\end{figure}

We conclude the presentation of results by discussing a selection of simulations using the Carman-Kozeny permeability function~\eqref{eq:CKPi} to examine the influence of the permeability variation on the flow dynamics. The stable solution branches of $F_s(\lambda)$ using the cubic permeability function~\eqref{eq:cubicPi} are compared in figure~\ref{fig:FluxPermComp}\textit{a} with those using the Carman-Kozeny permeability function~\eqref{eq:CKPi} in figure~\ref{fig:FluxPermComp}\textit{b}, with all other common parameters the same in both plots.
The two different permeability functions produce the same qualitative structure of flow dynamics, with the usual hysteresis loop involving a state of no flow for $\lambda<\lambda_u$ (blue line with triangles) and an upper branch of chimneys convection for $\lambda>\lambda_c$ (black solid curve). This indicates that the overall qualitative structure of the flow dynamics is robust for both of the different permeability functions. The main quantitative differences are that the Carman-Kozeny permeability permits a stronger flow with larger solute fluxes, and the bifurcation points $\lambda_c$ and $\lambda_u$ are at smaller $\lambda$ which indicates that chimney convection is more favourable for the Carman-Kozeny permeability than for the cubic permeability. 

\begin{figure}
  \psfrag{x}{$x$}
% \psfrag{z}{\hspace{-0.1cm} $z$}
% \psfrag{Phi}{\hspace{0.05cm} $\phi$}
% \psfrag{Th}{\hspace{0.05cm} $\theta$}
  \psfrag{z}{\hspace{-0.1cm} $z$}
 \psfrag{Phi}{\hspace{0.05cm} $\phi$}
 \psfrag{Th}{\hspace{0.05cm} $\theta$}
 \psfrag{eps}{$a$}
 \psfrag{a}{(\textit{a})}
  \psfrag{b}{(\textit{b})}
   \psfrag{c}{(\textit{c})}
    \psfrag{d}{(\textit{d})}
  \psfrag{e}{(\textit{e})}
   \psfrag{f}{(\textit{f})}
                                  \includegraphics[height=13.75cm,angle=270,trim= 0mm 30mm 0mm 23mm, clip=true]{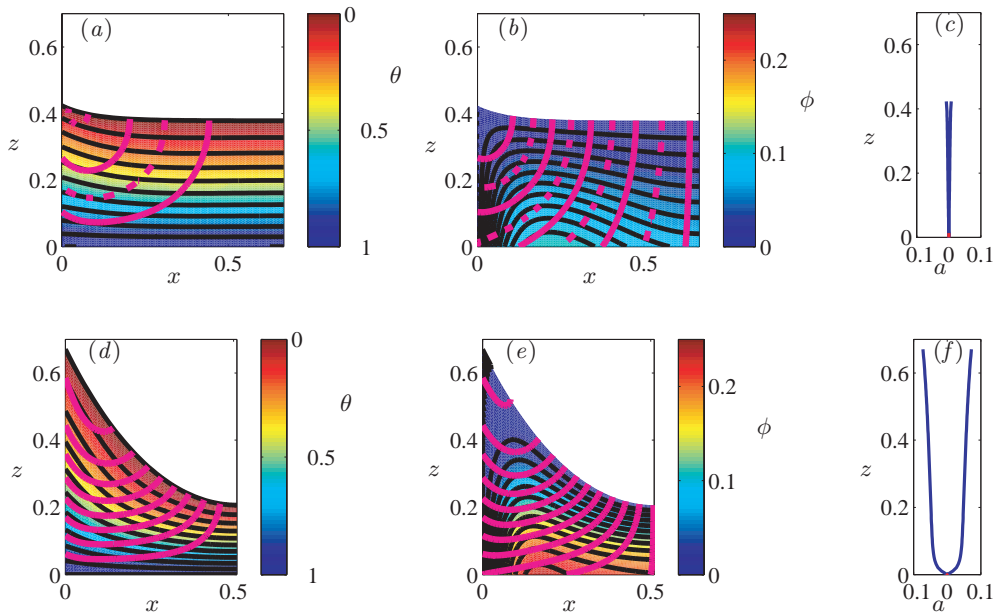}
\caption{Comparison of steady-state mushy layer profiles with optimal chimney spacing $\lambda=\lambda_o$ for cubic permeability function~\eqref{eq:cubicPi} (panels \textit{a}-\textit{c}) and Carman-Kozeny permeability function~\eqref{eq:CKPi}  (panels \textit{d}-\textit{f}). The colour scales show dimensionless temperature $\theta(x,z)$ in panels (\textit{a}, \textit{d}), and solid fraction $\phi(x,z)$ in panels (\textit{b}, \textit{e}) with black solid contours of constant $\theta$ and $\phi$ respectively. Streamlines of Darcy velocity $\mathbf{u}$ are shown by magenta curves with increment $\Delta\Psi=0.25$ in panels (\textit{a},\textit{d}), and streamlines of the net fluid flux $\mathbf{q}=\mathbf{u}-\mathbf{k}$ are illustrated in panels (\textit{b},\textit{e}). In addition, the dashed magenta curves in panels~(\textit{a},\textit{b}) show intermediate contours with increment $\Delta\Psi=0.125$. Panels (\textit{c}, \textit{f}) illustrate the variation of the chimney width $a(z)$ with height. All other parameters are identical to figure~\ref{fig:FluxPermComp}.}
\label{fig:KCcubicprofiles}
\end{figure}

The optimal solute flux also occurs at a smaller chimney half-spacing $\lambda_o$ for the Carman-Kozeny permeability and in figure~\ref{fig:KCcubicprofiles} we compare the corresponding flow profiles for cubic permeability (panels \textit{a}-\textit{c}) and Carman-Kozeny permeability (panels \textit{d}-\textit{f}). 
The solid fraction $\phi$ shows greater variation for the Carman-Kozeny permeability than the cubic permeability (see colour scales in figures~\ref{fig:KCcubicprofiles}\textit{e} and \textit{b} respectively), which will result in a greater contrast in permeability $\Pi(\phi)$ across the mushy region. This suggests flow focussing as a possible mechanism to explain the difference in optimal chimney spacing, with the larger permeability contrast for the Carman-Kozeny flow encouraging the flow to concentrate in narrower convection cells in order to reduce the resistance to flow. The narrower chimney spacing generates a larger baroclinic torque $\Rm\Pi\partial\theta/\partial x =O(\Rm/\lambda)$ in the vorticity equation~\eqref{eq:vorticity} which drives stronger flow. This is consistent with flow streamlines for the Darcy velocity $\mathbf{u}$ in figures~\ref{fig:KCcubicprofiles}\textit{a,d} and net fluid flux $\mathbf{q}=\mathbf{u}-\mathbf{k}$ in figures~\ref{fig:KCcubicprofiles}\textit{b,e}, both of which indicate stronger flow for the Carman-Kozeny permeability. The stronger flow enhances vertical heat transport through the chimney and results in the isotherms having steeper slope at the chimney wall consistent with the boundary condition~\eqref{eq:chimneyheatflux}. This results in the larger curvature of isotherms observed for the Carman-Kozeny permeability (figure~\ref{fig:KCcubicprofiles}\textit{d}) than for the cubic permeability (figure~\ref{fig:KCcubicprofiles}\textit{a}). The larger isotherm slope for the Carman-Kozeny permeability also contributes to increasing the baroclinic torque and driving stronger flow. Note that the mush--liquid interface is an isotherm, and also shows significant curvature for the Carman-Kozeny permeability in a manner reminiscent of previous enthalpy method simulations~\cite[compare to figure~10 of][]{Katz:2008tg}. The chimney is wider for the Carman-Kozeny permeability in order to accommodate the larger flux of fluid escaping the mushy layer.

\section{Conclusions \label{sec:discussion}}

We have considered finite-amplitude convection in a mushy layer with a periodic array of chimneys formed during directional solidification, and studied the parametric dependence of the flow over a wide range of chimney spacings, Rayleigh numbers and concentration ratios. A selection of simulations with a Carman-Kozeny permeability \eqref{eq:CKPi} illustrate that the flow dynamics is qualitatively similar to flow with a cubic permeability function~\eqref{eq:cubicPi}, with the Carman-Kozeny permeability resulting in stronger flow and narrower chimney spacings, potentially consistent with the process of flow focussing. For the cubic permeability function and a fixed concentration ratio, stable states of chimney convection exist for sufficiently large chimney spacings and sufficiently large Rayleigh number, consistent with a stability boundary~\eqref{eq:criticalRm} characterised by the effective Rayleigh number~\eqref{eq:modifiedRayleigh} based on the convection cell width $l/2$ when $lV/2\kappa\ll 1$. This recognizes the role of the horizontal wavelength in controlling mode restriction for cells that are narrower than the vertical thermal length scale. As the concentration ratio increases, the observed states of chimney convection are stable over a wider range of $\lambda$ and $\Rm$.

The optimal potential energy flux criterion of \cite{Wellsetal:2010} was  hypothesized as a physical mechanism to select preferred chimney spacings observed during directional solidification, and the scaling behaviour of these optimal flux states has been investigated here. For moderately large Rayleigh numbers $\Rm\gg1$ with $\Rm/\Cc=O(1)$, scaling laws~\eqref{eq:hlampsiscale}, \eqref{eq:fluxscale} and~\eqref{eq:phiscale} have been developed consistent with numerical simulations. These scalings illustrate that as the strength of convection increases, the flow velocity increases by generating a larger number of narrower convection cells of $O(1)$ aspect ratio in a thinner mushy region. The scalings derived here are also observed in a recent simplified model that assumes constant permeability and a linear background temperature gradient \citep{ReesJonesWorster:submit}, which provides support that their approximations do not have a major influence on the flow scaling. The flow dynamics in the chimney exhibits two asymptotic regimes: buoyancy and pressure gradients imposed from the neighbouring mush are of similar magnitude for $\Rm^2\Da \ll 1$ producing the scaling~\eqref{eq:chimscale1}, whilst for $\Rm^2\Da \gg 1$ the chimney forms an entirely passive conduit that adjusts in width to accommodate the fluid flux from the mushy region, giving the scaling~\eqref{eq:chimscale2}. For $\Rm/\Cc=O(1)$ the scaling laws recover a solute flux that increases linearly with Rayleigh number, consistent with the results of~\cite{Wellsetal:2010}, \cite{Wellsetal:2011}, and a recent approximate model \citep{ReesJonesWorster:submit}. However, the linear scaling does not persist indefinitely and for $\Rm\gg\Cc$ we observe sub-linear growth resulting from saturation of the flow due to porosity-dependent permeability as the solid fraction increases. This flow saturation for $\Rm\gg\Cc$ raises interesting questions regarding the applications in sea ice modelling, where typically $\Cc\ll1$. Intriguingly, \cite{Wellsetal:2011} found that the salinity evolution in experimental sea ice growth was consistent with a  scaling $F_s\sim\alpha( \Rm-\Rmo)$, despite being in the regime $\Cc\ll1$. It remains an open question to determine whether directional solidification will again recover $F_s\propto \Rm$ in a new dynamical regime for $\Cc\ll1$, or whether this is associated with some specific property of transient solidification where the ice thickness $h$ increases over time. A further question of interest relates to how convection in the neighbouring liquid region, which was neglected in this study, interacts with the mushy layer flow dynamics described here. If there is turbulent convection in the liquid, one might intuitively expect a thermal and compositional boundary layer to develop ahead of the mush--liquid interface, so that the interfacial temperature and concentration differ from those of the far-field liquid. One would also need to account for modified heat and solute fluxes at the mush--liquid interface, and the precise details of this interaction remain to be explored. Nevertheless, it is hoped that the scaling behaviour and system properties illustrated here may provide insight for simplified modelling applicable to both sea ice growth and other problems involving mushy-layer convection.

\paragraph{}
The authors appreciate the detailed comments and criticisms of the referees, D. Rees Jones, and M. G. Worster.  They thank C. Doering for discussions at the beginning of this project, and Yale University under the Bateman endowment for support of this research and JSW thanks the Wenner-Gren and John Simon Guggenheim Foundations, and the Swedish Research Council.

 \begin{appendix}
 
 \section{Heat flux at the chimney wall\label{app:lubrication}}
 
 In this appendix we show that the boundary condition~\eqref{eq:chimneyheatflux} for the heat flux at the chimney wall, previously derived by~\cite{SchulzeWorster:1998} in the asymptotic limit $\Rm^{4/3}\Da\ll 1$, can be justified as a perturbation approximation under the \textit{ansatz} $a/h\ll 1$ and $\psi=O(1)$ for a wider range of $\Rm$ and $\Da$. Motivated by this ansatz, we use rescaled coordinates $\xi=x/\epsilon h(0,t)$, $\zeta=z/h(0,t)$ to resolve flow within the chimney, and a rescaled chimney width $\mathcal{A}=a/\epsilon h(0,t)$. It is assumed that $\xi$, $\zeta$ and $\mathcal{A}$ are all of order one and $\epsilon\ll 1$, so that the chimney occupies $0\leq\xi\leq \mathcal{A}$ and $a/h\ll 1$. If $\psi=O(1)$, then in rescaled co-ordinates the heat equation~\eqref{eq:heat} within the chimney reduces to the leading-order balance
 \begin{equation} \frac{\partial^2 \theta}{\partial \xi^2} =O(\epsilon).\label{eq:rescalechimneyheat} \end{equation}
 Integrating twice and applying the symmetry condition $\partial \theta/\partial \xi=0$ at $\xi=0$ yields $\theta=f(\zeta)+O(\epsilon)$ for some function $f(\zeta)$, so that $\theta $ is independent of $\xi$ at leading order. 
 
To determine the heat flux at the chimney wall, we consider the first order correction to~\eqref{eq:rescalechimneyheat}, with~\eqref{eq:heat} yielding
 \begin{equation} \frac{\partial^2 \theta}{\partial \xi^2} =\epsilon\frac{\partial\psi}{\partial \xi} \frac{\partial \theta}{\partial \zeta}+ O(\epsilon^2), \label{eq:rescalechimneyheatnextorder} \end{equation}
 where we have taken advantage of the fact that $\partial \theta/\partial \xi=O(\epsilon)$. Noting that $\partial \theta/\partial \zeta$ is independent of $\xi$ at leading order, integrating~\eqref{eq:rescalechimneyheatnextorder} once and using the boundary conditions $\partial \theta/\partial \xi=\psi=0$ at $\xi=0$ yields the condition
 \begin{equation} \frac{\partial \theta}{\partial \xi} =\epsilon\psi \frac{\partial \theta}{\partial \zeta}+ O(\epsilon^2), \label{eq:rescalechimneyheatflux} \end{equation}
 which is valid for $0\leq\xi\leq \mathcal{A}$. Evaluating~\eqref{eq:rescalechimneyheatflux} at $\xi=\mathcal{A}$, and then returning to $(x,z)$ co-ordinates recovers the heat flux condition~\eqref{eq:chimneyheatflux} at the chimney wall. 
 
 \section{Notes on the numerical procedure\label{app:numerics}}

Two complementary numerical methods were applied, as described in the main text. We elaborate on the time-dependent code in \S\ref{app:timedep} and the arc-length continuation scheme in \S\ref{app:arclength}. 

\subsection{Time-dependent solution \label{app:timedep}}

The time-dependent problem was solved using finite differencing. We use the co-ordinate transform $\zeta=z/h$ to allow the use of a time-independent spatial grid with $0\leq \zeta \leq 1$.  Semi-implicit Crank-Nicholson timestepping is used to update the enthalpy $\theta -\phi \St/\Cc$ and the total concentration $\Cc \phi+\theta(1-\phi)$, using streamfunction data from the previous timestep only. The tridiagonal system resulting from~\eqref{eq:salt} was solved using standard LAPACK routines~\citep{laug}. After time discretization, \eqref{eq:heat} yields an elliptic system in space which, along with the vorticity equation~\eqref{eq:vorticity}, was solved using multigrid iteration via an adaptation of the MUDPACK multigrid routines~\citep{Adams:1989}. The boundary condition~\eqref{eq:chimneypsi} for the vorticity equation for the updated streamfunction $\psi^{n+1}$ is split in the form
\begin{equation} \psi^{n+1}=(1-\epsilon_1\,\Delta t) \psi^n + \epsilon_1\Delta t \left\{\left[\frac{a^3}{3\Da(1-\phi)^3}+a\right]\frac{\partial \psi}{\partial x} + \frac{3}{20} \frac{\Rm}{\Da}a^3 (\theta+1)\right\}^n , \end{equation} 
where $\epsilon_1$ is constant and $\Delta t$ the numerical time step. This relaxation approach satisfies~\eqref{eq:chimneypsi} with a discretization error of $O(\Delta t)$.  Using streamfunction data $\psi^n$ from the previous timestep allows a straightforward implementation of the MUDPACK multigrid routines when $a=0$ over a subsection of the boundary. The modified free-boundary condition~\eqref{eq:lidupdate} and~\eqref{eq:hshape} is solved by minimizing the least-squares residual from the relation
\begin{equation} \partiald{}{h_o} \int_0^{\lambda} \left\{ \nvec \cdot \nabla \theta+\theta_{\infty} \left[ \left(\uvec - \mathbf{k} \right) \cdot \nvec - \nabla \cdot \nvec \right] \right\}^2 \, dx =0, \end{equation}
which gives an algebraic equation for a new estimate $h_o=h^{*}$ at each timestep. The updated values of $h_o$ and $a(z,t)$ are then calculated using relaxation with
\begin{equation} \partiald{h_o}{t}=\epsilon_2 \left(h_o-h^{*}\right), \qquad \mbox{and} \qquad \partiald{a}{t}=\epsilon_3 \left(\partiald{\theta}{t}-\partiald{\theta}{z}+\uvec \cdot \nabla \theta\right). \end{equation}
The choice of the relaxation rates $\epsilon_1$, $\epsilon_2$, and $\epsilon_3$ used here influence the solution behaviour over very short timescales, but do not influence the final steady state except for close approach to the bifurcation points, where arc-length continuation was used to confirm the solution trajectory. Values $\epsilon_1\leq30$, $\epsilon_2\leq20$, and $\epsilon_3\leq 5$ were found to give good numerical convergence, with the maximal values typically used to obtain rapid relaxation compared to the development of the flow. Each subsection of the computer code was tested independently for accurate convergence against analytic solutions of simple advection and diffusion problems, and the full code was tested by adding false forcing to generate known analytic solutions of the modified system. The final solutions were tested by reducing step size $\Delta x$, and checking for $O(\Delta x^2)$ convergence.

\subsection{Arc-length continuation \label{app:arclength}}

A pseudo-arclength continuation method was used to trace the branches of stable and unstable steady states~\citep{Keller:1977}. Using $N_x\times N_x$ spatial grid points, the finite-difference solution has on the order of $N=3 N_x^2-4N_x+2$ degrees of freedom accounting for the variables $\theta(x,z)$, $\phi(x,z)$, $\psi(x,z)$ and $a(z)$, along with $h_0$ and $h_1$ which characterize $h(x,t)$, and subtracting known $\theta$, $\phi$ and $\psi$ values specified at boundaries. Rather than evaluating and performing inversions for the resulting large dense $N \times N$ Jacobian matrix, in an analogue with pseudo-spectral methods we seek a lower order representation of the solution expanded in Chebyshev polynomials
\begin{equation} (\theta,\phi,\psi)=\sum_{n,m=0}^{M,M} (\theta_{mn},\phi_{mn},\psi_{mn}) T_m\left(2\frac{x}{\lambda}-1\right) T_n(2\zeta-1), \qquad a(z)=\sum_{n=0}^{M} a_n T_n(2\zeta-1), \label{eq:chebexpand}\end{equation}
where the Chebyshev polynomials are conveniently defined by
\begin{equation} T_n(\cos{\theta})=\cos{n\theta} \qquad \mbox{for} \quad 0\leq \theta \leq \pi. \label{eq:chebdef} \end{equation}
As a result of the identity~\eqref{eq:chebdef}, the finite-difference solution is straightforwardly projected onto the Chebyshev polynomials using Fast Fourier Transforms.
 
The pseudo-arclength continuation scheme, which is an application of the method of~\cite{Keller:1977}, proceeds as follows. We symbolically represent the finite difference version of~\eqref{eq:heat}--\eqref{eq:symmetrybcs} as
\begin{equation} \frac{\partial \mathbf{y}}{\partial t} = f(\mathbf{y},\gamma) ,\label{eq:symbolic} \end{equation}
where $\mathbf{y}=\{\theta,\phi,\psi,a,h_0,h_1\}$ represents the finite-difference solution and $\gamma$ is the parameter to be varied. We let $\mathcal{P}$ denote projections into Chebyshev space according to \eqref{eq:chebexpand} with $\mathcal{P}^{-1}$ the corresponding inversion. Two initial steady states $\mathbf{y}_0$ and $\mathbf{y}_{-1}$, with $f(\mathbf{y}_0,\gamma_0)=f(\mathbf{y}_{-1},\gamma_{-1})=0$, form an initial approximation to the tangent vector in $(\mathcal{P}\mathbf{y},\gamma)$ space viz., 
\begin{equation} \mathbf{t}_0 = \frac{\left(\mathcal{P}\mathbf{y}_0-\mathcal{P}\mathbf{y}_{-1},\gamma_0-\gamma_{-1}\right)}{\| \left(\mathcal{P}\mathbf{y}_0-\mathcal{P}\mathbf{y}_{-1},\gamma_0-\gamma_{-1}\right)\|}. \end{equation}
We seek a new solution with pseudo-arclength step $\Delta s$ by using the predictor step $(\mathcal{P}\mathbf{y}_1,\gamma_1)=(\mathcal{P}\mathbf{y}_0,\gamma_0)+\Delta s \mathbf{t}_0$, and a corrector based on Newton iteration to solve the coupled system
\begin{align} \mathcal{P}f(\mathbf{y}_0,\gamma_0)&=0, \\
 \mathbf{t}_0 \cdot (\mathcal{P}\mathbf{y}-\mathcal{P}\mathbf{y}_0,\gamma-\gamma_0)&=\Delta s, \label{eq:pseudoarc} \end{align}
for the new solution $(\mathcal{P}\mathbf{y},\gamma)$ in Chebyshev space. Note that all the necessary function evaluations of $\mathcal{P}f(\mathbf{y},\gamma)$ are carried out sequentially as follows.  First, we transfer the Chebyshev projection $\mathcal{P}\mathbf{y}$ back to finite difference space using $\mathbf{y}=\mathcal{P}^{-1}\mathcal{P}\mathbf{y}$.  Secondly, we compute the time evolution of~\eqref{eq:heat}--\eqref{eq:symmetrybcs} with $a(z)$ held fixed to estimate $f(\mathbf{y},\gamma)$.  Finally, we return to the Chebyshev space via $\mathcal{P}f(\mathbf{y},\gamma)$ with the residual in satisfying~\eqref{eq:chimneymargeq} used in place of estimates of $\partial a/\partial t$.  After the corrector step yields a new solution estimate $(\mathcal{P}\mathbf{y},\gamma)$, a steady version of the finite-difference code is used to post-process the solution. Iteratively repeating the above predictor-corrector process allows the trajectory of steady-state solution branches to be traced.

\end{appendix}

\bibliographystyle{jfm}
\bibliography{WellsetalJFMMushBib}

\end{document}